\documentclass{elsart}

\usepackage{graphicx}

\usepackage{amssymb}

\begin{document}

\begin{frontmatter}



\title{Nuclear Structure, Random Interactions and Mesoscopic
Physics}


\author[nscl]{Vladimir Zelevinsky}
\author[fsu]{Alexander Volya}
\address[nscl]{National Superconducting Cyclotron Laboratory and
Department of Physics and Astronomy, Michigan State
University, East Lansing, MI 48824-1321, USA}
\address[fsu]{Department of Physics, Florida State University, Tallahassee,
FL 32306-4350, USA}

\begin{abstract}
Standard concepts of nuclear physics explaining the systematics of
ground state spins in nuclei by the presence of specific coherent
terms in the nucleon-nucleon interaction were put in doubt by the
observation that these systematics can be reproduced with high
probability by randomly chosen rotationally invariant
interactions. We review the recent development in this area, along
with new original results of the authors. The self-organizing role
of geometry in a finite mesoscopic system explains the main
observed features in terms of the created mean field and
correlations that are considered in analogy to the random phase
approximation.
\end{abstract}

\end{frontmatter}

\section{Introduction}

It became a common place to claim that the basic facts of the
nuclear ground state systematics, namely that all even-even nuclei
have ground state spin $J_{0}=0$ and the lowest possible isospin
$T_{0}=T_{{\rm min}}$, are due to the fundamental properties of
residual nucleon-nucleon forces. The {\sl pairing} phenomenon was
known, in particular through the mass formula, from the beginning
of nuclear spectroscopy. It was formulated by Racah in an elegant
form with the use of the seniority quasispin formalism
\cite{Racah,Talmi}. The predictions of the simple shell model by
Mayer and Jensen \cite{MJ} would be uncertain for all nuclei,
except magic ones and those with one particle or one hole on top
of the magic core, if the pairing would not allow one to guess
that the nuclear ground state spin in odd-$A$ nuclei is determined
by the last unpaired nucleon. A. Bohr, Mottelson and Pines
\cite{BMP} found a profound analogy of nuclear pairing to
superconducting pair correlations of electrons in metals, and
Belyaev  in his seminal paper \cite{Bel} demonstrated how the
pairing interaction creates correlations that modify not only the
ground state energy but all single-particle and collective
properties of low-lying nuclear states, in a good agreement with
observed trends \cite{KS}.

Following the advances in experiments and theory towards nuclei
far from stability, the interest to pairing has recently
increased. The pairing correlations play a decisive role in
determining the nuclear drip-line; many nuclides can be stable
only due to the pairing correlations between the outermost loosely
bound nucleons. The pairing interaction is an inalienable and very
important part of all modern shell model versions \cite{BAB}.

Therefore an unexpected observation by Johnson, Bertsch and Dean
\cite{JBD} was met with great interest and immediately put under
the microscope of various tests. They noticed that even {\sl
randomly} taken two-body forces acting between the fermions in a
restricted Hilbert space of few single-particle orbitals lead to
the statistical predominance of the ground state spin $J_{0}=0$.
The broad discussion that followed this discovery revealed that
similar phenomena take place for interacting bosons as well. The
natural questions arise: do we understand well the physics
generated by random interactions under constraints of rotational
symmetry and what is the reason for empirical regularities in
nuclei? The problem is not limited to nuclear physics. Atomic
clusters, particles in the traps, quantum dots and disordered
systems, such as quantum spin glasses, are just a few examples
where the same questions are to be answered.

From a more general theoretical viewpoint related to {\sl
many-body quantum chaos} \cite{grib,big,ann}, we deal with closed
mesoscopic systems that generically display chaotization of motion
due to intrinsic interactions. In the absence of a heat bath and
external disorder, the interactions play the role of a randomizing
or thermalizing factor. They create a very complicated structure
of the eigenstates. However, the presence of exact symmetries
(rotational, time-reversal, parity, isospin) leads to
non-vanishing correlations between the classes of states with
different exact quantum numbers since they are governed by a
single deterministic (even if randomly picked) Hamiltonian with a
relatively small number of parameters. Such correlations bring a
new, hardly discussed before, element to theory of quantum chaos.

Below we describe the problem more in detail following the main
ideas proposed for explanation by various authors. We show that a
conventional notion of a mean field created by the interaction
removes the main puzzling features of the problem and puts the
whole story on a clear track. Of course, the open questions still
remain leaving the room for future exciting studies.

\section{Two-body interactions in an isolated many-body system
and many-body quantum chaos}
\subsection{Hamiltonian}
Our starting point is a standard {\sl shell-model approach} to a
many-body problem. $N$ particles are interacting through two-body
forces within Hilbert space that is built on a certain number of
single-particle orbitals. We label the single-particle states as
$|1)$, incorporating all necessary quantum numbers in the unified
label {\sl 1}. Using the single-particle basis diagonalizing the
independent particle part, the general Hamiltonian of the system
can be written as
\begin{equation}
H=\sum_{1}\epsilon_{1}a^{\dagger}_{1}a_{1}+\frac{1}{4} \sum_{1234}
V(12;34)a^{\dagger}_{1}a^{\dagger}_{2}a_{3}a_{4},  \label{2.1}
\end{equation}
where we introduced the creation and annihilation operators with
usual commutation (anticommutation) rules for bosons (fermions).
The interaction matrix elements $V(12;34)$ are correspondingly
symmetrized (antisymmetrized) with respect to permutations
$1\leftrightarrow 2$ and $3\leftrightarrow 4$. Assuming
time-reversal invariance, we can consider all matrix elements
real. The restriction to two-body forces (``rank" of the
interaction $r=2$) is not significant as long as the total
particle number $N\gg r$, and can be removed. Note that the
general form (\ref{2.1}) does not explicitly carry any
conservation law except for the particle number. Later we consider
the requirements of rotational (or isospin) invariance.

Two physical formulations can be considered in parallel. In
application to a realistic system, the Hamiltonian is derived from
more general theory (for example, for nuclei it can be based on
meson theory or quark models) or built empirically with the
parameters, $\epsilon_{1}$ and $V(12;34)$, adjusted to
experimental data. One can also consider {\sl ensembles} of
Hamiltonians that satisfy the requirements of Hermiticity and
quantum statistics but the parameters, or part of them, are
treated as random variables taken from some distribution. The
explicitly introduced randomness of the Hamiltonian keeping the
same form as that of actual mesoscopic systems was used with the
purpose to bring the {\sl global} description in terms of random
matrices near to physical reality. It turns out that the {\sl
local} spectral statistics, starting from sufficiently high level
density, are universal. They express generic properties of
many-body quantum chaos (plus the assumption of time reversal
invariance) and do not depend on the details of the interaction.
Below we first characterize these universal features and then turn
to the ground state problem.

\subsection{Ensemble of random interactions}

The studies of the random two-body interactions go back to
Wigner-Dyson {\sl random matrix theory}, see \cite{Fluc} and
references therein. This stage of development was thoroughly
reviewed by Brody {\sl et al.} in Ref. \cite{Brody}; see also the
latest review \cite{Guhr}. The {\sl two-body random ensemble},
TBRE \cite{TBRE,TBREa}, in contrast to full canonical (Gaussian
Orthogonal, GOE, or Gaussian Unitary, GUE) ensembles \cite{Mehta},
considers matrices in many-body Hilbert space, where nonzero
off-diagonal matrix elements link the independent particle states
that can be connected by two-body processes but not constrained by
any conservation laws, except for the symmetry dictated by the
particle statistics. These matrix elements are taken in the TBRE
as uncorrelated and normally distributed real random quantities.
More general {\sl embedded ensembles} \cite{Kota} can be
considered with $r$-body forces \cite{embe} for $r<N$; the case
$r=N$ with a simultaneous interaction of all particles returns to
the full GOE. The angular momentum conservation was, as a rule,
ignored because of severe mathematical difficulties \cite{Brody}.

We see essential new properties of the TBRE as compared to the
canonical random matrix ensembles. (i) The orthogonal, or unitary,
invariance of the statistical distribution of random matrix
elements is lost. (ii) The natural basis is that of independent
particle configurations where only configurations that differ by
not more than the occupancies of a pair of orbitals in the initial
and final states can be connected by a single-step interaction. In
this basis the many-body Hamiltonian matrix is sparse. (iii) The
nonzero matrix elements of this matrix are strongly correlated.
Indeed, a given two-body scattering process may occur on the
background of many different spectator configurations of remaining
particles; all many-body matrix elements in those cases are equal
regardless of a random or deterministic character of the two-body
interaction.

Before the work \cite{JBD}, the studies of the TBRE did not
consider the consequences of rotational invariance so that the
single-particle levels did not carry any additional quantum
numbers being fully characterized by their energy. The most
important conclusion of these studies was the {\sl ``chaotic"}
character of local spectral statistics, essentially the same as
predicted for the GOE in spite of a very different distribution of
many-body matrix elements. Many results are insensitive to the
exact form of the distribution function of random two-body matrix
elements that can differ from the Gaussian still remaining
symmetric with respect to the sign. It was established \cite{mon}
that the global (``secular") behavior of the level density in the
given finite Hilbert space of a certain number of single-particle
orbitals is close to Gaussian for $N>r$ while for $N=r$ it tends
to the semicircle typical for the GOE or GUE. There are only few
analytical results for the TBRE and its modifications \cite{Kota},
although significant numerical work has been done.

\subsection{Complexity of many-body states}

The detailed shell model studies for complex atoms \cite{grib} and
nuclei \cite{big} showed that realistic forces, Coulomb for atoms
and semiempirical effective nucleon-nucleon interactions for
nuclei, generate the local spectral statistics well described by
the GOE and TBRE within each class of many-body states with fixed
exact quantum numbers. Considering the dependence on the
interaction strength, the chaotic statistics of nearest level
spacings and the so-called $\Delta_{3}$ statistics of level number
fluctuations emerge when the interparticle interactions are turned
on with their strength still much weaker than the realistic value.
This happens without any randomness in the Hamiltonian, in spite
of correlations due to the two-body character of the forces and
the fact that the realistic distributions of the many-body matrix
elements are generically close to exponential rather than Gaussian
\cite{big}. The mechanism of spectral chaotization is provided by
{\sl multiple avoided crossings} of levels inside a fixed symmetry
class.

As the interaction strength increases beyond threshold for onset
of spectral chaos, and level dynamics with less frequent crossings
loses its turbulent character, the main ongoing process is the
growth of complexity of the eigenfunctions. The important question
here is how one can quantify the {\sl degree of complexity} of an
{\sl individual} wave function. The specification of a wave
function is always related to a certain basis $|k\rangle$. In the
eigenbasis of the Hamiltonian, each eigenfunction has just one
component that obviously indicates the absence of complexity. In
the above mentioned process of switching on interaction, it is
natural to refer all eigenstates to the original basis of
noninteracting particles and follow the gradual increase of
complexity measured by the number of significant components in the
wave function. This choice of the reference basis is also singled
out by many-body physics. In a realistic system of the type we are
interested in, the single-particle structure is determined by the
self-consistent field due to all particles. The mean field
embodies the most regular effects of the interaction. The residual
interactions already do not contain such average components.
Therefore one can think of the mean field basis as the best choice
for {\sl separating} regular and chaotic aspects brought in by the
interaction \cite{mf}.

The degree of complexity of an eigenstate $|\alpha\rangle$ with
respect to the reference basis $|k\rangle$ can be quantified with
the help of Shannon {\sl information entropy} \cite{izr,entr,big}.
If the eigenfunction is given by the normalized superposition
\begin{equation}
|\alpha\rangle=\sum_{k}C^{\alpha}_{k}|k\rangle,     \label{2.2}
\end{equation}
information entropy of the state $|\alpha\rangle$ in the basis
$|k\rangle$ is defined in terms of the weights $w^{\alpha}_{k}=
|C^{\alpha}_{k}|^{2}$ as
\begin{equation}
I_{\alpha}=-\sum_{k}w^{\alpha}_{k}\ln w^{\alpha}_{k}.  \label{2.3}
\end{equation}
As the interaction strength increases, information entropy grows
from zero in principle being able to reach the limit of $I_{{\rm
max}}=\ln d$, where $d$ is the space dimension. This maximum
possible value can be realized for the fully delocalized function
with all equal weights $w^{\alpha}_{k}=1/d$. In the GOE the
average value of $I_{\alpha}$ is lower than this limit, $I_{{\rm
GOE}}=\ln (0.48 d)$, because of the requirements of orthogonality
of different eigenstates.

The shell model analysis \cite{entr,big} shows that information
entropy in all symmetry classes grows smoothly with the interaction
strength and in the middle of the spectrum gets close to the
$I_{{\rm GOE}}$. With the interaction strength artificially
increased beyond its realistic value, one can reach the GOE limit
uniformly in excitation energy \cite{temp,big}. It is important
that for the realistic, and therefore consistent with the mean
field, interaction strength information entropy $I_{\alpha}$ is a
smooth monotonously increasing to the middle of the spectrum
function of excitation energy $E_{\alpha}$. This allows one to
treat information entropy as a {\sl thermodynamic variable} and
build up the corresponding temperature scale \cite{temp,big,ann}
avoiding any reference to a heat bath or Gibbs ensemble. Thus, one
can consider thermodynamics of a closed mesoscopic system based on
typical properties of individual quantum states.

The physical foundation for that is given by the chaotic mixing of
states as a result of the strong interaction at a high level
density. This mixing makes statistical properties of closely
located states {\sl uniform} (thereby the question by Percival
\cite{perc} on a generic relation between the complicated
neighboring states is solved - ``the states look the same") and
guarantees that macroscopic observables do not depend on the exact
population of adjacent microscopic states and the corresponding
phase relationships. This is exactly what is needed for the
statistical description. Such considerations shed new light on a
problem of justification of the thermodynamic approach for closed
mesoscopic systems.

The quantity $d_{\alpha}=\exp(I_{\alpha})/0.48$ can be interpreted
as an effective number of significant (``principal") components of
the wave function, or its localization length. The components
$C_{k}^{\alpha}$ of a complicated wave function $|\alpha\rangle$
on average are uniformly distributed over a sphere of dimension
$d_{\alpha}$. The fully uniform distribution on a $d$-dimensional
sphere is restricted only by the normalization,
\begin{equation}
P_{d}(C_{1},...,C_{d})=\frac{\Gamma(d/2)}{\pi^{d/2}}\,
\delta(\sum_{k=1}^{d}w_{k}-1),                \label{2.4}
\end{equation}
which leads to the distribution function of any chosen component
$C$
\begin{equation}
P_{1}(C)=\frac{\Gamma(d/2)}{\sqrt{\pi}\Gamma((d-1)/2)}\,
(1-w)^{(d-3)/2},                               \label{2.5}
\end{equation}
where, as in eq. (\ref{2.4}), $w=C^{2}$. In the asymptotic limit
of large $d$, this distribution goes to the Gaussian. The square
of the amplitude has a distribution
\begin{equation}
P_{1}(w)=\frac{\Gamma(d/2)}{\sqrt{\pi}\Gamma((d-1)/2)}\,
\frac{1}{\sqrt{w}}\,(1-w)^{(d-3)/2}             \label{2.6}
\end{equation}
that goes to the Porter-Thomas ($\chi^{2}$) distribution for large
$d$. The realistic distributions of the components in the nuclear
shell model \cite{verbaar,big}, except for the lowest and the
highest states, are close to these predictions with local values
of $d_{\alpha}$ smoothly changing along the spectrum, The strength
distribution (\ref{2.6}) along with the nearest level spacing
distribution can serve as an experimental means for recovering the
strength missing in the background of experiments that cannot
resolve the invisible fine structure \cite{kilg}.

The complexity measure $I_{\alpha}$ or $d_{\alpha}$ gives a tool
for estimating matrix elements of simple operators between a
simple and complicated state or between two complicated states. In
both cases, the typical reduction of the matrix element compared
to that between two simple (let say, noninteracting) states is
given by the factor $1/\sqrt{d}$ if one assumes that the two
complex states have a similar degree of complexity. Since the
corresponding level density, which determines energy denominators,
increases on average $\propto d$, we come to the {\sl statistical
enhancement} of perturbations, $\propto \sqrt{d}$, in the region
of many-body quantum chaos \cite{SF}. In light nuclei this
enhancement can be seen directly in shell model calculations
\cite{AB}. Remarkable examples are given by the strong enhancement
of weak interactions in nuclear neutron resonances (parity
violation in polarized neutron scattering \cite{Alfim,Bowm,Mitch}
and fragment asymmetry in fission by polarized neutrons
\cite{Dan,Petr}). Here again we see that statistical regularities
in a mesoscopic system coexist with the opportunity to reveal,
both theoretically and experimentally, properties of individual
quantum states.

\subsection{Chaos and thermalization}

 Another aspect of the same problem is the possibility to
describe complicated eigenstates in the standard statistical
language of single-particle occupation numbers. It was noticed,
both for atoms \cite{grib}, and nuclei \cite{temp,big,ann}, that
expectation values of the occupation numbers $n_{j}$ of the mean
field orbitals are close to what would be predicted by Fermi-Dirac
statistics. Effective temperatures $T_{\alpha}$ extracted for
individual states $|\alpha\rangle$ are in good correspondence with
thermodynamic temperature determined by the level density as well
as with the information temperature found from Shannon entropy.
This shows that one can successfully use the notions of {\sl
Fermi-liquid theory} modeling the system as a gas of
quasiparticles not only near the ground state, as it is usually
assumed (in nuclei just in this region the description has to be
modified because of pairing correlations \cite{ZVYad}), but
practically at any excitation energy below decay threshold. The
finite lifetime of quasiparticles is simply translated into
statistical occupation factors different from 0 and 1 and smoothly
changing along the spectrum. The analytical description of the
process of equilibration was given by Flambaum, Izrailev and
Casati \cite{FIC}, and Flambaum and Izrailev \cite{FI} in the
framework of the TBRE.

We need again to stress that information entropy is capable to
characterize the degree of complexity only relative to a reference
basis. This can be considered as an advantage of information
entropy as a measuring tool since we are able to discover
relations between the eigenbasis and {\sl various} reference
choices. The special role of the {\sl self-consistent} mean field
basis is now seen in the consideration of the occupation numbers
defined with respect to this basis that forms a skeleton
supporting all complications induced by the interaction. The
equilibrating factor is the interparticle interaction rather than
a heat bath. For the interaction of rigid spheres, which is known
to generate chaotic dynamics, it was shown rigorously \cite{mark}
that the equilibrium momentum distribution is that of Boltzmann,
Bose-Einstein or Fermi-Dirac depending on the statistics of
particles even if the interaction cannot be reduced to gaseous
rare collisions. In realistic cases the interaction strength is in
accordance with the parameters of the mean field. In the case of
artificially enhanced interaction, all states go to the GOE limit
of complexity, and the single-particle thermometer is not capable
of resolving the spectral evolution \cite{temp,big}.

The description with the aid of information entropy does not take
into account any phase correlations between the components of an
eigenfunction. In a sense it gives a {\sl delocalization} measure
\cite{big,Kota} of the given state  in the original basis of
noninteracting particles. It cannot distinguish between an
incoherently mixed chaotic state and collective state that is a
regular superposition of many basis states with certain phase
relationships. The information approach may be also inadequate for
an unstable mean field or a phase transition occurring at some
temperature (excitation energy). Here another way of
characterizing the individual quantum states may be useful
\cite{Sok}. One can look at the response of a given state
$|\alpha\rangle$ to external noise described by random parameters
$\lambda$ in the Hamiltonian. The averaging over $\lambda$
determines the density matrix
\begin{equation}
\rho^{\alpha}_{kk'}=\langle C^{\alpha}_{k}(\lambda)
C^{\alpha\ast}_{k'}(\lambda)\rangle_{{\rm av}}        \label{2.7}
\end{equation}
that allows one to define von Neumann entropy
\begin{equation}
S_{\alpha}=- {\rm Tr}\{\rho^{\alpha}\ln(\rho^{\alpha})\}.
                                                 \label{2.8}
\end{equation}
In contradistinction to information entropy, this quantity, that
may be called {\sl invariant correlational entropy} (ICE), does
not depend on the choice of representation and takes into account
correlations between the amplitudes of the wave function.

The ICE is very sensitive to quantum phase transitions. If the
random parameter $\lambda$ fluctuates around the phase transition
point, the strong variation of the structure of the state gives
rise to a peak in the ICE as was shown in the interacting boson
model (IBM) \cite{cej} and in the realistic shell model
\cite{ICE}. One can notice also a common physical aspects shared
by the ICE and the notion of {\sl fidelity} extensively studied
recently in considerations of quantum dynamics related to quantum
echo, decoherence, Zeno effect and quantum computing, see for
example \cite{Peres,Niels,Pros}.

\section{Rotational invariance}
\subsection{Role of symmetries}
From the very beginning of studies of random matrices and quantum
chaos, the crucial influence of {\sl global symmetries} was
repeatedly stressed by many authors. Random matrix ensembles make
averaging over all Hamiltonians in a given universality class
\cite{Fluc}. The classes are fully determined by the fundamental
symmetries as Hermiticity, both for GOE and GUE, and time-reversal
invariance (for GOE). The additional requirement for the canonical
Gaussian ensembles is the invariance of the distribution of matrix
elements under orthogonal or unitary basis transformations. This
last demand expresses the limiting property of extreme chaos and
brings at our disposal the necessary reference point, against
which we can look at the realistic systems with their specific
deviations from this limit. The ensembles as TBRE do not obey this
requirement but this does not influence the local spectral
statistics. In addition an exact permutational symmetry for
fermions or bosons is also imposed here.

Self-sustaining mesoscopic systems reveal other exact symmetries,
first of all rotational symmetry (in the absence of external
fields). As a result, any eigenstate is a member of a degenerate
rotational multiplet $|JM\rangle$ with total spin $J$ and its
projection on the laboratory quantization axis $J_{z}=M$. The
classes of states with different quantum numbers of $J$ and $M$
are not mixed, and one
can study the onset of chaos, spectral characteristics, complexity
of wave functions and so on for each class separately. The
situation is similar, for example, to the Sinai billiard, where
the studies responsible for a hypothesis \cite{boh} of the
correspondence between quantum level statistics and classical
chaos can be tested using one octant of the billiard and
continuing wave functions to the entire area according to the
symmetry class. Combining states of various classes into a common
spectrum, one comes to the Poissonian level statistics
\cite{GuP,Brody}. To the best of our knowledge, correlations
between the states of different exact symmetry in the same
billiard were not studied.

At the same time, serious efforts were applied to the problems of
approximate symmetries, onset of chaos along with destruction of
symmetry, transition from the GOE to GUE due to violation of
time-reversal invariance by the magnetic field or $T$-odd nuclear
forces, intermediate spectral statistics and so on \cite{Guhr}.
The example most relevant to nuclear structure is given by the
{\sl isospin invariance} \cite{HRM}. The classes of nuclear states
with different isospin are mixed by electromagnetic interactions
and strong forces violating charge symmetry, and this can be seen
in transition probabilities and reaction amplitudes. In the shell
model versions with exact isospin conservation and without weak
interactions, the classes of states are characterized by exact
quantum numbers $J^{\pi}T$.

The common Hamiltonian that governs nuclear dynamics certainly
establishes correlations between the states of different classes
even if they belong to the region of quantum chaos. One can
imagine, for instance, a deformed system with extremely chaotic
many-body dynamics inside. Nevertheless, the rotational invariance
guarantees the existence of the rotational branch of the
excitation spectrum with energy $E_{J}$ at least approximately
given by $AJ(J+1)$. In a macroscopic system, this would be a
continuous Goldstone mode that emerges because of the spontaneous
orientational symmetry breaking by the choice of the body-fixed
frame; in a finite system it is simply a rotational band, or, for
a chaotic system, ``compound band" as suggested by Mottelson
\cite{Doss}. Thus, we obtain a clear correlation between the
states of the same band in different $J$-classes. The intrinsic
structures of these states should also be close. In a sense, this
might even be a classical rigid body with microscopic quantum
chaos of interacting constituents.

\subsection{Geometric chaoticity}

Consider a finite many-body system with exact angular momentum
conservation. The total spin of the system,
\begin{equation}
{\bf J}=\sum_{a} {\bf j}_{a},                  \label{3.2}
\end{equation}
is built up of spins of individual constituents. As a number of
particles grows, so does the number of independent ways of
building a many-body state of a given total spin $J$. This number
determines the dimension of a given $J$-class, $d_{J}$; a similar
construction is necessary for a total isospin $T$.  Those
independent combinations correspond to various recouplings of
spins in the process of constructing the full state:
\begin{equation}
\{[(j_{1}j_{2})j_{12}j_{3}]j_{123}j_{4}\}j_{1234}...J.
                                                 \label{3.3}
\end{equation}
Different paths to the same values of $J$ can be distinguished by
high $nj$-symbols or coefficients of fractional parentage.

In the shell model, even with a particle number of the order of
10, the number of different paths, eq. (\ref{3.3}), is large, and
the resulting products of many consecutive Clebsch-Gordan
coefficients (CGC) determine the orthogonal combinations within
the same class. Let us, for example, look at the class $J=0$ in a
system of an even particle number. We can start with a simple
state of seniority $s=0$, when we couple particles pairwise to
$j_{12}=j_{34}=...0$ (to guarantee the full permutational symmetry
it is more convenient to use the pair operators in secondary
quantization, as we do below). Here all CGC are trivial, and the
state has a very regular structure. Each new state should be
orthogonal to all previous ones, so that at some point we have to
employ another combination (a closed loop or few loops of vectors
${\bf j}_{a}$) that proceeds through different intermediate
stages. At a sufficiently large dimension, the majority of paths
look as a {\sl random walk process of vector coupling}. This
source of randomness we call geometrical chaoticity.

The property of geometrical chaoticity was practically used long
ago by Bethe \cite{Bethe}, see also \cite{Eric}, to derive the
partial nuclear level density $\rho(E;J)$ for a given angular
momentum in the model of noninteracting fermion gas. Assuming that
the projections $j_{z}=m$ of particle spins are coupled into the
total projection $M$ in a random walk process, one can apply the
central limit theorem and come to the Gaussian probability $w(M)$
of a given value of $M$ with zero mean and the variance
\begin{equation}
(\Delta M)^{2}=N\langle m^{2}\rangle,            \label{3.4}
\end{equation}
expressed in terms of the (energy-dependent) number $N$ of active
fermions (particles and holes) and average single-particle value
of $m^{2}$ in space of available orbitals. The level density with
given $M$ is then given by $\rho(E;M)=\rho(E)w(M)$, where
$\rho(E)$ is the total level density. Using the standard trick,
one can get the approximate expression for the multiplicity
$d_{J}$ of states with given spin $J$,
\begin{equation}
\frac{d_{J}}{d}=w(J)-w(J+1)\approx \frac{2J+1}{2\sqrt{2\pi}
(\Delta M)^{3}}e^{-J(J+1)/2(\Delta M)^{2}}.     \label{3.5}
\end{equation}
This result, invalid for the largest values of $J$, shows that the
maximum of the multiplicity is near $J=\Delta M-1/2$. From eq.
(\ref{3.4}) we see that it grows $\propto \sqrt{N}$.

Regrettably, it is very hard to develop a statistical theory for a
random process with quantized vectors as ${\bf j}_{a}$ that are
coupled not algebraically. This would be equivalent to developing
statistical theory of fractional parentage coefficients. In spite
of various attempts and  some useful results that can be extracted
from works by Wigner \cite{Wig}, Ponzano and Regge \cite{regge}
and others \cite{Bieden,Wong}, such theory still does not exist.

The reality of geometric chaoticity is clearly seen in the nuclear
shell model \cite{big}. Prior to any diagonalization, one can
derive important characteristics of the energy spectrum directly
from the Hamiltonian matrix. A basis state $|JT;k\rangle$ of
independent particles {\sl projected} onto certain values of spin
and isospin is a superposition of the stationary states
$|JT;\alpha\rangle$ with the same real amplitudes $C^{\alpha}_{k}$
that determined the composition of the eigenstate in eq.
(\ref{2.2}). The weights $w^{\alpha}_{k}$ define the {\sl strength
function} $F_{k}(E)$ of the simple state $|JT;k\rangle$ according
to
\begin{equation}
F_{k}(E)=\sum_{\alpha}w^{\alpha}_{k}\delta (E-E_{\alpha}),
                                                   \label{3.6}
\end{equation}
where $E_{\alpha}$ are energies of the eigenstates. The strength
function (since $\sum_{k}F_{k}(E)=\rho(E)$, it is called {\sl
local density of states} in condensed matter theory, where the
basis states $|k\rangle$ are localized ones) determines the time
evolution of the state $|k\rangle$ prepared at the initial moment.
The centroid of the strength function $F_{k}(E)$ is given by the
{\sl diagonal} matrix element of the Hamiltonian, $H_{kk}$. The
energy dispersion $\sigma_{k}$ of the state $|JT;k\rangle$ can be
found as
\begin{equation}
\sigma^2_{k}=\langle k|(H-H_{kk})^{2}|k\rangle=\sum_{l\neq k}
H_{kl}^{2},                                    \label{3.7}
\end{equation}
the sum of all {\sl off-diagonal} matrix elements in the $k^{{\rm
th}}$ row of the Hamiltonian matrix. A remarkable fact is that in
a given shell model class $(JT)$ the dispersions $\sigma_{k}$ are
nearly constant, $\sigma_{k}\approx \bar{\sigma}$, for all states
$|JT;k\rangle$. This equilibration for {\sl noninteracting}
particles comes only from the $JT$-projection, and therefore is
the direct output of geometric chaoticity that accompanied the
projection algorithm. Parenthetically we can mention that the
constant magnitude of the dispersion (\ref{3.7}) is important for
determination of the {\sl spreading width} of the strength
function \cite{Fraz} that approaches the limit of $2\bar{\sigma}$
in the case of strong fragmentation.

\subsection{Rotationally invariant two-body Hamiltonian}

Now we explicitly introduce the requirements of rotational
invariance in the Hamiltonian (\ref{2.1}), both for
single-particle states and the interaction. We assume spherical
symmetry of the mean field with orbitals characterized by the
angular momentum $j$, its projection $j_{z}=m$, and isospin
projection $\tau_{3}$. For definitiveness we consider fermions
with isospin 1/2 and assume that every value of $j$ appears only
once. The Hamiltonian of the system is determined by the set of
single-particle energies $\epsilon_{j}$ that, under conditions of
rotational and isospin invariance, do not depend on $m$ and
$\tau_{3}$, and by the interaction that preserves the total
angular momentum $L$ and total isospin $t$ of the interacting
pair. The eigenstates of the Hamiltonian have exact quantum
numbers of total angular momentum, $J$ and $J_{z}=M$, and total
isospin, $T$ and $T_{3}$.

\begin{figure}
\begin{center}
\includegraphics[width=7 cm]{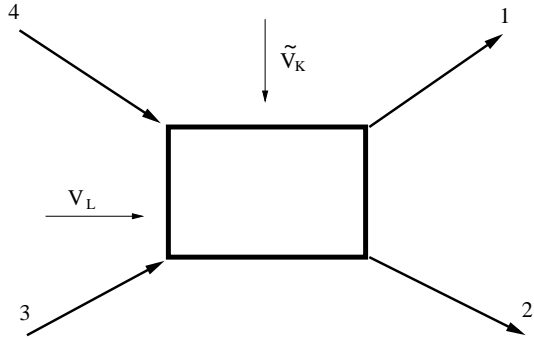}
\end{center}
\caption{\label{diagram}The diagram of the two-body interaction in
particle-particle and particle-hole channels characterized by the
total spins, $L$ and $K$, respectively.}
\end{figure}

The most general form of the two-body interaction under such
conditions is
\begin{equation}
H_{int}=\sum_{L\Lambda,tt_{3};\{j\}}V_{Lt}(j_{1}j_{2};j_{3}j_{4})
P^{\dagger}_{L\Lambda,tt_{3}}(j_{1}j_{2})P_{L\Lambda,tt_{3}}(j_{3}j_{4}).
                                                      \label{3.8}
\end{equation}
Here we use the pair annihilation and creation operators for each
$Lt$-pair channel,
\[P_{L\Lambda,tt_{3}}(j_{1}j_{2})=\frac{1}{\sqrt{1+\delta_{j_{1}j_{2}}}}
[a_{j_{1}}a_{j_{2}}]_{L\Lambda,tt_{3}},\]
\begin{equation}
P^{\dagger}_{L\Lambda,tt_{3}}(j_{1}j_{2})=\frac{1}
{\sqrt{1+\delta_{j_{1}j_{2}}}}
[a^{\dagger}_{j_{2}}a^{\dagger}_{j_{1}}]_{L\Lambda,tt_{3}},
                                                  \label{3.9}
\end{equation}
where the vector coupling with the appropriate CGC to the total
rotational ($L\Lambda$) and isospin ($tt_{3}$) quantum numbers of
the pair is implied, Fig. \ref{diagram}. One can also remove isospin
invariance taking different values for proton and neutron
single-particle levels and interaction matrix elements but still
preserving angular momentum conservation.

In this formulation, the numerical parameters of the Hamiltonian
are, apart from the single-particle energies, the interaction
amplitudes $V_{Lt}$ that do not depend on projections $\Lambda$
and $t_{3}$. The number $k$ of these amplitudes is typically much
smaller than the number $d_{JT}$ of levels in the $JT$-class (the
classes with the largest possible values of $J$ and $T$ might be
exceptional in this respect). Thus, the well studied nuclear $sd$
shell model includes $d_{5/2}, s_{1/2}$ and $d_{3/2}$ orbitals for
neutrons and protons. This model is completely defined by three
single-particle energies and 63 interaction matrix elements
$V_{Lt}$ while, for example, with 12 valence fermions, there are
839 states in the class with lowest quantum numbers $JT=00$
\cite{big}, the largest class in this model appears with $JT=31$
and has $d_{JT}=6706$.

It is always possible to recouple the creation and annihilation
operators from the particle-particle channel used in eqs.
(\ref{2.1}) and (\ref{3.8}) to the particle-hole channel, see Fig.
\ref{diagram}, where the Hamiltonian would have the multipole-multipole
structure $\{[a^{\dagger}a]_{K} [a^{\dagger}a]_{K}\}_{00}$ with
the spin $K$ of the particle-hole pairs; the similar recoupling is
performed for isospins. The corrections to the orbital energies
(one-body terms) appear in this process. The transformation can be
done in two ways which correspond to the crossing transition from
the $s$-channel to $t$- and $u$-channels in quantum field theory.
It is important to stress that, although not seen immediately, the
number of parameters is the same in different channels because of
the permutational symmetry of identical particles, as will be
shown better in the next subsection.

The typical class dimensions of few hundred up to few thousand
make the nuclear shell model a perfect tool for studying the exact
solution of a many-body problem. With dimensions of this size, the
diagonalization is straightforward and we obtain all individual
eigenfunctions while the statistical regularities are already
clearly pronounced. Thus, we are in a typical realm of mesoscopic
physics. Again we stress that {\sl all} classes of states are
described by the same matrix elements and therefore have to be
correlated.

The particle-particle interaction channel with $L=0$ and $t=1$,
when $j_{1}=j_{2}$ and $j_{3}=j_{4}$, represents conventional
isospin-invariant isovector pairing. This component of the
interaction is, as a rule, considered to be responsible for the
main pairing effects in binding energy. It is possible that around
the $N=Z$ line the isoscalar pairing, $t=0$, is important,
especially in nuclei far from stability; for classification of
various types of pairing see, for example \cite{good} and
references therein. According to the generalized Pauli principle
for a two-nucleon system, the symmetries in spin-spatial variables
and in isospin are complementary. The symmetry relation for the
pair operators (\ref{3.9}) reads
\begin{equation}
P_{L\Lambda,tt_{3}}(j_{1}j_{2})=(-)^{j_{1}+j_{2}+L+t}
P_{L\Lambda,tt_{3}}(j_{2}j_{1}).                      \label{3.10}
\end{equation}
In particular, for $j_{1}=j_{2}$, only even values $L=0,2,...,
2j-1$ are allowed for $t=1$ (or for identical fermions without
isospin) while $t=0$ requires the odd values $L=1,3,...,2j$. With
time reversal invariance, the parameters $V_{Lt}$ can be chosen
real.

\subsection{Single $j$-level}

Here we give the formalism for the simplest fermionic space,
namely that of $\Omega=2j+1$ single-particle states $|jm)$ of a
single $j$-level for one kind of particles. With this
simplification, it will be possible to see the core of the
problem. In this case the mean field part is just a constant
proportional to the particle number, and the Hamiltonian takes the
form
\begin{equation}
H=\epsilon N+\sum_{L}V_{L}\sum_{\Lambda}P^{\dagger}_{L\Lambda}
P_{L\Lambda},                                      \label{3.11}
\end{equation}
where the pair operators are defined in terms of the $3j$-symbols
as
\begin{equation}
P^{\dagger}_{L\Lambda}=\frac{1}{\sqrt{2}}\sum_{m_{1}m_{2}}
\sqrt{2L+1}(-)^{L-\Lambda}\left(\begin{array}{ccc}
                       j & L & j\\
                   m_{1} & -\Lambda & m_{2}\end{array}\right)
a^{\dagger}_{1}a^{\dagger}_{2}.                    \label{3.12}
\end{equation}
{\sl Only even} $L$ pairs are present in eq. (\ref{3.11}) so that
the number of independent interaction parameters $V_{L}$ is
$k=j+1/2$; all unnecessary labels are omitted.

We define the multipole operators in the particle-hole channel,
Fig. \ref{diagram}, as
\begin{equation}
M_{K\kappa}=\sum_{m_{1}m_{2}}(-)^{j-m_{1}}\left(\begin{array}{ccc}
    j & K & j\\
-m_{1} & \kappa & m_{2}\end{array}\right)a^{\dagger}_{2}a_{1}.
                                                \label{3.13}
\end{equation}
Here any integer value of $K$ from 0 to $2j$ is allowed.  The
Hermitian conjugation gives
\begin{equation}
M^{\dagger}_{K\kappa}=(-)^{\kappa}M_{K-\kappa}.     \label{3.14}
\end{equation}
The special important cases are the particle number operator
\begin{equation}
N=\sum_{m}a^{\dagger}_{m}a_{m}=\sqrt{\Omega}\,M_{00},
                                                 \label{3.14a}
\end{equation}
and the angular momentum (in spherical components)
\begin{equation}
J_{\kappa}=\sum_{mm'}(m|j_{\kappa}|m')a^{\dagger}_{m}a_{m'}=
\sqrt{j(j+1)\Omega}\,M_{1\kappa}.                 \label{3.14b}
\end{equation}

The alternative, particle-hole, form of the Hamiltonian is
\begin{equation}
H=\tilde{\epsilon}N-\frac{1}{2}\sum_{K}\tilde{V}_{K}\sum_{\kappa}
M^{\dagger}_{K\kappa}M_{K\kappa},               \label{3.15}
\end{equation}
where the single-particle energy is renormalized,
\begin{equation}
\tilde{\epsilon}=\epsilon +
\frac{1}{2\Omega}\sum_{K}\tilde{V}_{K},            \label{3.16}
\end{equation}
and the interaction parameters are transformed according to
\begin{equation}
\tilde{V}_{K}=(2K+1)\sum_{L}(2L+1)\left\{\begin{array}{ccc}
                  j & j & L\\
                  j & j & K\end{array}\right\}V_{L}. \label{3.17}
\end{equation}
Inversely,
\begin{equation}
V_{L}=\sum_{K}\left\{\begin{array}{ccc}
                  j & j & K\\
j & j & L\end{array}\right\}\tilde{V}_{K}.       \label{3.18}
\end{equation}

Since a reversible algebraic transformation cannot increase the
number of independent parameters, there should exist constraints
that reduce the number of independent constants $\tilde{V}_{K}$
from $2j+1$ to the number $k=j+1/2$ of the original parameters
$V_{L}$. Indeed, the particle-hole amplitudes $\tilde{V}_{K}$ are
{\sl interrelated} through
\begin{equation}
\tilde{V}_{K}=(2K+1)\sum_{K'}(-)^{K+K'}\left\{\begin{array}{ccc}
                  j & j & K\\
j & j & K'\end{array}\right\}\tilde{V}_{K'}.       \label{3.19}
\end{equation}
These relations expressing the symmetry of recoupling between the
$t$- and $u$-channels are important for the dynamics of the model.
They were discussed more in detail by one of the authors
\cite{Volya}. In particular, we note the results for the pairing
interaction, $L=0$,
\begin{equation}
V_{0}=-\frac{1}{\Omega}\sum_{K}(-)^{K}\tilde{V}_{K}, \label{3.20}
\end{equation}
and monopole interaction, $K=0$,
\begin{equation}
\tilde{V}_{0}=-\frac{1}{\Omega}\sum_{K}\tilde{V}_{K}. \label{3.21}
\end{equation}

To conclude this section, we write down the commutator algebra of
the pair and multipole operators (we use the abbreviation $g_{K}
=\sqrt{2K+1}$),
\begin{equation}
[P_{L'\Lambda'},P^{\dagger}_{L\Lambda}]=\delta_{LL'}
\delta_{\Lambda\Lambda'}+2\sum_{K\kappa}g_{K}^{2}
 X^{LL';K}_{\Lambda \Lambda'\kappa}M^{\dagger}_{K\kappa};
                                                   \label{3.22}
\end{equation}
\begin{equation}
[P^{\dagger}_{L\Lambda},M_{K\kappa}]=2\sum_{L'\Lambda'}
X^{LL';K}_{\Lambda \Lambda'\kappa} P^{\dagger}_{L'\Lambda'};
                                                  \label{3.23}
\end{equation}
\begin{equation}
[M^{\dagger}_{K\kappa},M_{K'\kappa'}]=\sum_{S\sigma}[1-(-)^{K+K'+S}]
\frac{g_{S}^{2}}{\sqrt{g_{K}g_{K'}}}
X^{KK';S}_{\kappa\kappa'\sigma} M^{\dagger}_{S\sigma}.
                                                    \label{3.24}
\end{equation}
We introduced here the common geometric factor
\begin{equation}
X^{LL':K}_{\Lambda\Lambda'\kappa}=g_{L}g_{L'}
\left\{\begin{array}{ccc}
  L & L'& K\\
  j & j & j\end{array}\right\}(-)^{\Lambda}
  \left(\begin{array}{ccc}
  L & L' & K\\
  -\Lambda & \Lambda' & \kappa\end{array}\right).   \label{3.25}
\end{equation}
In the right hand side of eq. (\ref{3.22}), we did not indicate
explicitly the symmetry factors $\Theta_{L}=[1+(-)^{L}]/2$ and
$\Theta_{L'}$; similarly eq. (\ref{3.23}) contains $\Theta_{L}$.

The closed algebra of particle-particle and particle-hole
operators is too complicated for a general analysis. However, it
contains closed subalgebras with simpler properties. The operators
$P_{00},\,P^{\dagger}_{00}$ and $M_{00}\propto N$ form the well
known from seniority theory \cite{Racah,Talmi} {\sl quasispin
algebra} isomorphic to $SU(2)$ and widely used for the solution of
the pairing problem, approximate in BCS theory \cite{Bel} or exact
\cite{EP}. The odd-$K$ multipoles $M_{K\kappa}$ form the algebra
$U(2j+1)$, and the three components of angular momentum
proportional to $M_{1\kappa}$ give a standard $SU(2)$. The
algebraic properties of the operators will be used in the
equations of motion.

\section{Ordered spectra from random interactions?}

\subsection{Main evidence}

As mentioned in Introduction, the shell model calculations with
random two-body interactions (\ref{3.8}) provide unexpected
results. Using the $sd$- and $pf$-shell model for even-even nuclei
with various particle numbers and an ensemble of random parameters
$V_{Lt}$, the authors of Ref. \cite{JBD} found a surprisingly
large fraction of cases with the ground state spin $J_{0}=0$.
Being later confirmed and studied by many authors for different
ensembles and different fermion and boson single-particle spaces,
the effect seems to be generic, see for example
\cite{JBDT,BFP,BF,Bij,KZC,Kus,KSJ,Mul,Hor,oddA,Mul1,Cov,Yad,Mul2,Zhao,Zhao1,Zhao2,Arima,Yoshi,Zhao3,Droz,vela,Zhaocomm,velah,KAPP,KapPJ,Chau}.

The naive idea of what should be the ground state spin of a system
governed by random rotationally invariant interactions comes from
a simple counting of multiplicities $d_{J}(N)$ of states with a
given spin $J$ and particle number $N$ in Hilbert many-body space
spanned by a given set of single-particle orbitals. As stated in
Subsection 3.2, as a result of a random spin coupling, the spin
value with the largest multiplicity increases with the particle
number $\propto\sqrt{N}$. However, the empirical probabilities
$f_{J}$ of the ground state spin $J$ turn out to be very different
from a simple estimate $d_{J}/d$, where $d$ is the total dimension
of space. The original paper \cite{JBD} gives the following
results of direct repeated diagonalization and averaging over the
ensemble (we discuss the choice of the ensemble later on): for
$N=6$ identical particles in the $sd$-shell ($\Omega=12$) they
found $f_{0}=76\%$ and for the $pf$-shell ($\Omega=20$)
$f_{0}=75\%$, whereas the corresponding Hilbert space
multiplicities are equal to $d_{0}/d=9.8\%$ and 3.5\%,
respectively. Similar results were found for two kinds of
particles, the state with $J_{0}=T_{0}=0$ had a predominant
probability to be found as the ground state.

\begin{table}
$$
\begin{array}{|r||r|r|r|r|r|r|r|r|r|}
\hline
J& (a)  & (b)   & (c)   & (d)  & (e)  & (f) & (g) &(h)&(i)  \\
\hline
0&0.61&12.7 &  65.4 & 61.9  & 65.3 & 54.5 & 80.5 & 55.2 & 64.1 \\
2&1.45&3.6  &       &  0.8  &  0.8 &      &      & 0.6  &      \\
4&2.38&5.4  &  1.9  & 2.5   &  2.7 &      & 1.0  & 3.7  & 2.2  \\
5&2.15&1.8  &       &       &      & 9.1  &      &      &      \\
6&3.18&6.4  &  4.8  & 6.5   & 14.7 & 9.1  & 1.7  & 6.5  & 4.9  \\
8&3.74&3.6  &  3.4  & 2.6   &  2.2 &      & 1.8  & 4.7  & 3.3  \\
10&4.41&4.1 & 2.6  & 3.0   &  5.8 & 9.1  & 1.2  & 3.5  & 2.4  \\
12&4.53&4.4 &      &       &  1.3 &      &      & 1.0  &      \\
13&4.07&2.6 &     &       &      &      &      & 0.6  & 0.6  \\
16&4.49&3.0 &     &       &  0.7 &      &      & 0.7  &      \\
18&4.31&3.1 & 1.0 & 1.7   &      &      &      & 1.3  & 1.0  \\
28&2.05&1.4 & 0.9 & 1.2   &      & 9.1  &      & 0.9  & 1.0  \\
33&0.94&0.9 &     &       &      &      &      & 0.6  &      \\
36&0.66&0.9 &     & 0.7   &      &      &      &      &      \\
42&0.19&0.5 & 0.8 & 0.6   &      &      &      & 0.9  & 0.7  \\
46&0.05&0.3 & 0.8 & 1.0   &      &      &      & 0.8  & 0.7  \\
48&0.05&0.3 & 11.8& 11.5  &  1.4 & 9.1  & 9.2  & 12.4 & 11.7 \\
\hline
\end{array}
$$
\caption{\label{tab1}Statistics of ground state spins $J_{0}$ for
$N=6$ identical particles on a single $j=21/2$ orbital: $(a)$
multiplicity $d_{J}/d$; (b) predictions from approximating fermion
pairs with non-interacting bosons; $(c)$ fractions $f_{J}$ for a
random ensemble with the uniform distribution of all $V_{L}$ in
the interval [-1,1]; $(d)$ for a Gaussian random ensemble of
$V_{L}$ with zero mean and dispersion equal to 1; $(e)$ for the
uniform distribution of $V_{L}$ scaled by $(2L+1)^{-1}$; $(f)$
predictions according to a recipe of Ref. \cite{Zhao2}; $(g)$ for
the uniform ensemble of $V_{L\neq 0}$ and fixed attractive
pairing, $V_{0}=-1$; $(h)$ the same as in $(g)$ but with repulsive
pairing, $V_{0}=+1$; $(i)$ the same as in $(g)$ and $(h)$ but
without pairing, $V_{0}=0$. Except for columns $(a)$ and $(b)$,
only fractions $f_{J}>0.5\%$ are included. }
\end{table}

To display the universality of the effect we show in Table
\ref{tab1} the results of the diagonalization for various random
ensembles for a system of $N=6$ identical fermions on a single
level $j=21/2$. Comparing the column $(a)$ with $(c)$ and $(d)$, we see that
the approximately the same predominance of $J_{0}=0$ (it exceeds
the statistical multiplicity by an order of magnitude) occurs for
the uniform and Gaussian ensembles of the parameters $V_{L}$.
Moreover, even the suppression of high-$L$ components of the
interaction by a factor $(2L+1)^{-1}$ does not change $f_{0}$,
column $(e)$. Contrary to the statement of Ref. \cite{JBD} that
the choice of ensemble is crucial, we come to the conclusion
confirmed by other works that the predominance of $J_{0}=0$ is
insensitive to the specific features of the random ensemble.

The Gaussian ensemble used in \cite{JBD} assumed that the
interaction parameters are normally distributed uncorrelated
random variables with zero mean and the variance that has an extra
factor of 2 for the diagonal matrix elements. The latter property
was borrowed from canonical Gaussian ensembles and in reality does
not matter. An assumption was also made concerning the choice of
the variance $\overline{V_{L}^{2}}$ scaled as a function of $L$ as
$(2L+1)^{-1}$. According to the original idea of Ref. \cite{JBD},
this choice was made in order to obtain the most random
interaction with analogous statistical properties in the
particle-particle and particle-hole channels (``Random
Quasiparticle Ensemble", RQE). Regrettably, this idea cannot be
implemented since such a scaling is not invariant under the
transformation (\ref{3.17}) and (\ref{3.18}). The reason is in a
formally different number of parameters in the two channels that
is equalized by extra constraints on the parameters in the
particle-hole channel that are not independent, eq. (\ref{3.19}).
The non-equivalence of the channels could be easily seen if the
definition for the variance would be written in the way explicitly
including the factor $[1+(-)^{L}]/2$ necessary for Fermi
statistics, eq. (\ref{3.10}). Therefore, the RQE is just one of
possible choices without preferential meaning.

\begin{figure}
\begin{center}
\includegraphics[width=14 cm]{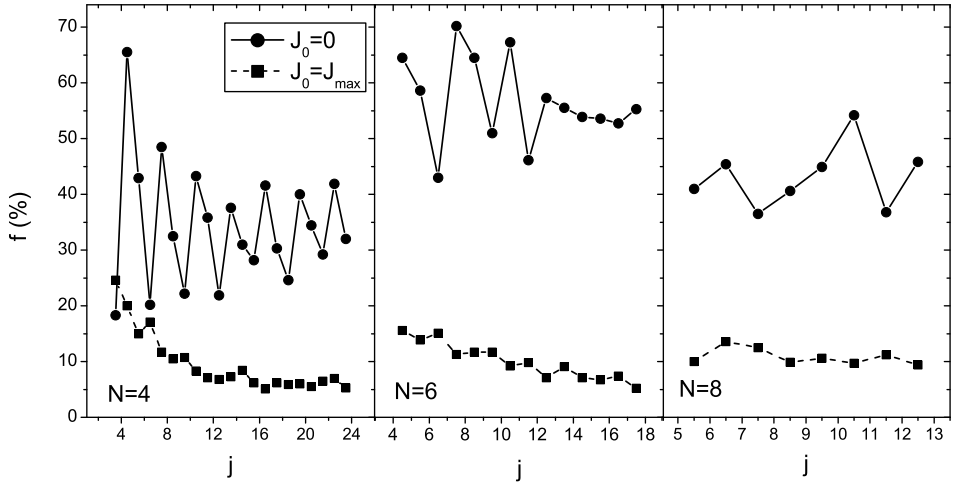}
\end{center}
\caption{Fractions of states with ground state spins $J_{0}=0$
(circles connected by solid lines) and $J_{0}=J_{{\rm max}}$
(squares connected by dashed lines) for the single$-j$ models as a
function of $j$. The three panels correspond to particle numbers
$N=4,6$ and 8, from left to right, respectively. The uniform
distribution of interaction parameters $V_{L}\in [-1,1]$ was used.
\label{pstat}}
\end{figure}

\begin{figure}
\begin{center}
\includegraphics[width=11 cm]{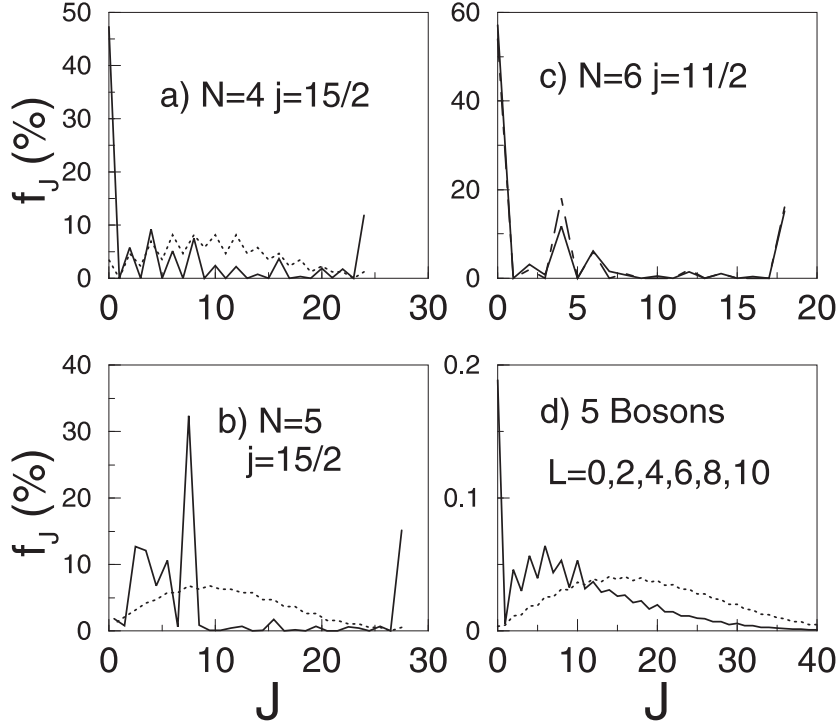}
\end{center}
\caption{\label{N4N5}Distributions $f_{J}$ of ground state spins:
$(a)$ 4 fermions on the $j=15/2$ level; $(b)$ 5 fermions on the
$j=15/2$ level; in this odd-$N$ system, in accordance to panel
$(a)$, the maximum of the probability corresponds to $J=j$; $(c)$
6 fermions on the $j=11/2$ level; $(d)$ 5 bosons with interaction
in the pair states with $L=0,2,4,6,8,10$, see Section 5. The
uniform distribution of $V_{L}$ was used. The dotted lines in
panels $(a),(b)$ and $(d)$ indicate the statistical distribution
of multiplicities $d_{J}/d$; the dashed line in panel $(c)$
corresponds to the case with no pairing, $V_{0}=0$.}
\end{figure}

Various assumptions concerning the choice of ensemble lead to
quite similar results for $f_{J}$ although details can differ.
Along with this, it was observed \cite{BF,Mul} that in many cases
the fraction of the ground state spins equal to the {\sl largest
possible|} spin $J_{{\rm max}}$ is also considerably enhanced, see
the last line of Table \ref{tab1}. Note that in the single-$j$
model (the results for $f_{0}$ and $f_{J_{{\rm max}}}$ in this
case are shown in Fig. \ref{pstat} as a fluctuating function of
$j$) the state with $J=J_{{\rm max}}$ is {\sl unique} being
constructed by full alignment of the particles along the
quantization axis (no random walk in this case). The effects
persist for odd-$A$ \cite{Mul}, see also Fig. \ref{N4N5}$b$, and
odd-odd \cite{oddA} nucleonic systems and for interacting bosons
\cite{BF}, see also Fig. \ref{N6N7}. Typical examples are shown in
Fig. \ref{N4N5}. Structures of the wave functions, properties of
the observables and excitation spectra also were studied in a
multitude of models. We will discuss the most important findings
as we go step by step testing the explanations put forward by
various authors.

\subsection{Induced pairing?}

The first idea suggested for the explanation of the ground state
spin effects \cite{JBD,JBDT} was related to Cooper-type pairing
induced by random interactions through some high-order mechanism.
For a single $j$-level there are deep symmetry reasons to presume
a special role of pairing since the seniority quantum number is
only broken by about $1/3$ of $k=j+1/2$ linearly independent
combinations of interaction parameters \cite{Volya,rowe}. The
pairing idea was supported \cite{JBD} by the enhanced gap between
the ground and first excited states in the case of $J_{0}=0$,
presence of odd-even staggering and the large pair transfer matrix
element $\langle N-2|P|n\rangle$ between the ground states of
adjacent even nuclei. The pair transfer operator $P$, however, was
taken in such a form (different in different realizations) that in
fact predetermined the result.

Table \ref{tab1} contains important information about the role of
explicit pairing (parameter $V_{0}$ in the Hamiltonian) in the
single-$j$ case. Eliminating pairing completely from the dynamics,
column $(i)$, does not noticeably change the fraction $f_{0}$.
Even the transition to ``antipairing" (fixed $V_{0}=+1$) only
slightly reduces the value of $f_{0}$, column $(h)$. The
attractive fixed pairing, $V_{0}=-1$, however, increases the
fraction $f_{0}$ to 80\%, column $(g)$. Similar results can be
seen in Fig. \ref{N4N5}$c$: in this case, $N=6$ and $j=11/2$,
fully random interactions lead to $f_0=55.9$\%, setting pairing to
zero $V_0=0$ reduces fraction of $J_0=0$ ground states to
$f_0=51.9$\%, in ``antipairing'' case $f_0=44.4$\% and finally,
forced pairing with $V_0=-1$ leads to $f_0=82.7$\%.

\begin{figure}
\begin{center}
\includegraphics[width=14 cm]{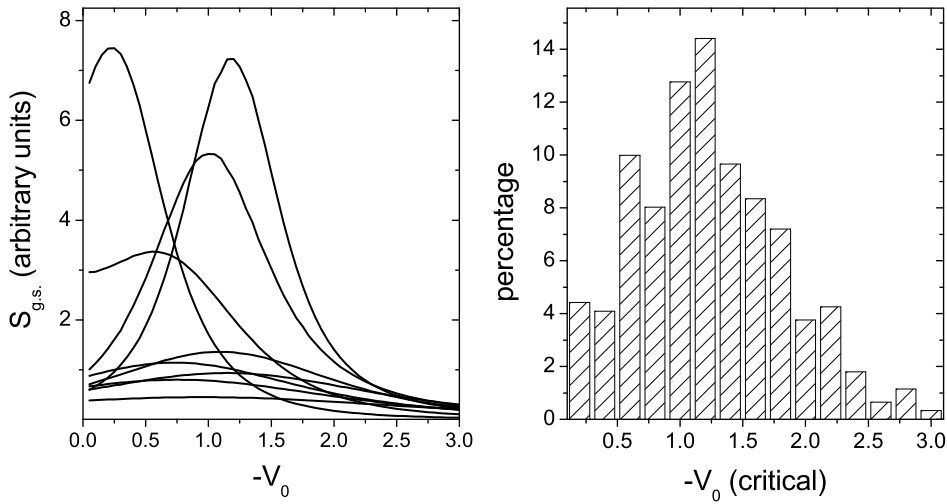}
\end{center}
\caption{\label{miscice} On the left panel the invariant correlational 
entropy (ICE)
for randomly selected realizations of ensemble is plotted
as a function of pairing strength $V_0$. The system of six
particles on $j=15/2$ level is chosen. Right panel shows the distribution of
critical values $V_0$ defined as peaks on each ICE curve.}
\end{figure}

The fact that, as a rule, the ground state with $J_{0}=0$ does not
contain considerable pairing correlations, can be seen from the
observation  based on the ICE, Section 2.4. Indeed, with forced
pairing (by means of large negative $V_0$), many random
realizations change the ground state structure undergoing a
transition to the superconducting paired state. The ICE can be
used here to study such transitions in each individual
realization. In Fig. \ref{miscice}, left, the behavior of ICE for a few
randomly selected realizations is shown as a function of the
pairing strength $V_0$, around which the fluctuations necessary to
obtain the ICE are imposed. All selected realizations had $J_0=0$
even without pairing, i.e. at $V_0=0.$ The peaks are observed in
the ICE curves for all random realizations. The location of the
peak roughly indicates the critical point of the phase transition
with the maximum sensitivity to noise, while the sharpness of the
peak and ICE magnitude reflect the size of the critical region. As
one would expect, the properties of the pairing phase transition
vary significantly from one sample to another. It is however clear
that, in order for a significant fraction of random realizations
to exhibit developed pairing, an average coherent attraction in
the $V_0$ channel must be added, see right 
panel in Fig. \ref{miscice}, an analog of a {\sl displaced}
ensemble \cite{vela,velah}.

\begin{figure}
\begin{center}
\includegraphics[width=14 cm]{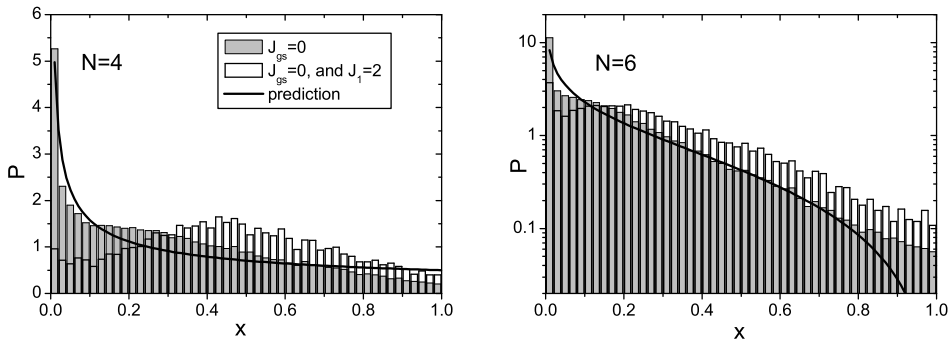}
\end{center}
\caption{\label{rpairing} The distributions of overlaps $x$
between the ground states of spin zero in the random ensemble and
states of seniority zero, for $N=4$, left, and $N=6$, right,
fermions on the $j=15/2$ level, shaded histograms, for the
ensemble of all random $V_{L}$. The solid lines show the
statistical distributions expected for a chaotic system of
dimensions $d=3$, left, and $d=4$, right. The unshaded
distributions are calculated for the parts of the ensemble that
give a ground state $J=0$ {\sl and} the first excited state
$J=2$.}
\end{figure}

To find out if there is an important role of induced pairing, we
compare directly the empirical ground state wave functions
$|0\rangle$ with $J_{0}=0$ from the diagonalization in the random
ensemble to the fully paired state $|s=0\rangle$ of seniority zero
and the same particle number that can be built uniquely for a
single-$j$ level. Fig. \ref{rpairing} shows with shaded histograms
the distribution $P(x)$ of the
overlaps
\begin{equation}
x=|\langle 0|s=0\rangle|^{2}, \quad 0\leq x\leq 1,  \label{4.1}
\end{equation}
for 4 and 6 fermions on the $j=15/2$ level, left and right panels,
respectively, that have $d_{0}=3$ and $d_{0}=4$ states of $J=0$.
The overlap (\ref{4.1}) is one of the weights $w$ of the wave
function, namely the one for the paired basis state. The paired
structure would give a peak of $P(x)$ at $x\rightarrow 1$ while a
random wave function is characterized by the distribution $P(x)$
of eq. (\ref{2.6}). For $d=3$, left panel,
\begin{equation}
P_{d=3}(x)=\frac{1}{2\sqrt{x}},                 \label{4.2}
\end{equation}
with a peak at $x\rightarrow 0$ shown with solid line. This
distribution appears in the problem of pion multiplicity from a
disordered chiral condensate \cite{dcs,ocs}, where isospin 1
determines the dimension $d=3$. Similarly, in the right panel,
with a bigger space, $d_{0}=4$, the chaotic distribution is
$P_{d=4}(x)\propto (1-x)^{3/2}x^{-1/2}$, and the empirical
distribution displays only a slight excess near $x=1$, of the
order of 1\% in the total normalization.

A similar analysis with similar results was also carried out for
the $sd$ shell model space \cite{Hor}, where the geometry is much
richer and includes also isospin. The calculation of information
entropy (\ref{2.3}) revealed the chaotic character of eigenstates;
even the ground state always has a high degree of complexity
exceeding that for the realistic shell model. One can compare the
ground state wave functions $|JT=00\rangle$ obtained with random
interactions to those corresponding to the realistic system. With
the standard effective interaction for this shell \cite{BW} (and
the results were shown to change only slightly for different
interactions), the comparison for $^{24}$Mg ($N=8$ nucleons)
provides the following average overlaps $x$ with the ground state
wave functions for different random ensembles: (a) degenerate
single-particle energies and all 63 two-body matrix elements
generated as random uniformly distributed quantities:
$f_{0}=0.591,\, x=0.020$; (b) realistic single-particle energies
and random interaction matrix elements: $f_{0}=0.493,\,x=0.053$;
(c) realistic single-particle energies and six pairing interaction
matrix elements with $L=0$ and $t=1$, and random remaining 57
matrix elements: $f_{0}=0.678,\,x=0.106$; (d) single-particle
energies and 57 non-pairing matrix elements set to zero, while the
pairing matrix elements taken as random variables:
$f_{0}=0.922,\,x=0.052$. In the cases of realistic single-particle
levels and random interactions, the ensembles of two-body
amplitudes were chosen in such a way that have a realistic ratio
of their magnitude to the single-particle level spacings
\cite{Hor} in order to allow for fair comparison.

Although the large value of $f_{0}$ is common for all variants, we
see again that the presence of regular pairing, case (c),
increases the fraction $f_{0}$. The largest $f_{0}$ is observed in
case (d), where the off-diagonal pair transfer amplitudes make
quantum numbers $J=T=0$ preferable for an even number of pairs,
whereas the competing influence of incoherent interactions and the
mean field level splitting is absent. The average overlaps $x$ are
small in confirmation of the conclusion carried over from the
single-$j$ model that the ground state wave functions with random
interactions are far away from realistic ones which are up to high
extent determined by pairing, although presence of deformation in
realistic nuclei is another factor reducing average $x$.

In systems with many double degenerate orbitals for spins $j=1/2$
\cite{KapPJ}, the fraction $f_{0}$ turns out to be close to 100\%,
which, at least partly, is, like in the previous case (d), induced
by a great preponderance of off-diagonal pair transfers. Here one
should mention that such a system reminds a quantum spin glass
with random spin-spin interactions. In that case \cite{Sush} the
ground state spin increases $\propto\sqrt{N}$ as expected for
random spin coupling. The crucial difference as compared to
shell-model systems is in the type of random coupling. For quantum
spin glasses the spin-spin interaction $({\bf s}_{1}\cdot{\bf
s}_{2})$ is usually assumed. This is equivalent to a fixed
relation 3:1 between the singlet and triplet parts of the
interaction. Contrary to that, in shell-model systems those parts
are fully uncorrelated.

One can notice that even in case (a) the average overlap of 2\% is
higher than what we would expect, eq. (2.6), from the uniform
distribution of the components, $\sim 1/d\sim 0.3\%$ for the
actual dimension $d=325$. The maximum effect of 11\% is reached in
case (c) due to the combination of two effects. First, the
presence of realistic pairing lowers energies of states with
paired particles. Second, the effective dimension $d^{\ast}$ is
now smaller than $d$ because the contributions of non-paired
states to the ground state are appreciably reduced. The
stabilizing presence of the mean-field orbitals, case (b), also
increases the overlap with the shell model ground state. To
conclude, a small effect of induced pairing should be present but
not as a main reason for the statistical predominance of $f_{0}$.

\subsection{Time-reversal invariance?}

The normal pairing ($Lt=01$ channel) is believed to be singled out
as the most important part of residual nucleon-nucleon forces due
to the maximum overlap of spatial wave functions of the paired
particles \cite{BMP}. The coherent effects of this residual
attraction are enhanced by a greater density of states for such
pairs since in this case any single-particle state $|1)$ is
coupled to its time-reversed counterpart $|\tilde{1})$ that has,
in the absence of external magnetic or Coriolis fields, exactly
the same single-particle energy (Kramers degeneracy). The same
reasoning is behind the assigning a special role to the Cooper
pairs with zero total momentum and singlet spin state in
superconductors.

Since the dynamical specificity of pairing forces converts the
ground state into a condensate of time-reversed pairs, it is
natural to invert the problem and hypothesize that the
time-reversal invariance can imply the preponderance, at least in
the statistical sense, of the ground states with $J_{0}=0$
\cite{vela}. This idea was checked by Bijker, Frank and Pittel
\cite{BFP} who explicitly included in the random two-body
interaction a Hermitian but not $T$-invariant part assuming the
interaction Hamiltonian in the form that was earlier studied as an
example of the transition from the GOE to the GUE,
\begin{equation}
H=\cos\alpha\,H^{R}+i\sin\alpha\,H^{I},           \label{4.4}
\end{equation}
where $H^{R}$ and $H^{I}$ are uncorrelated Gaussian variables with
zero mean and the same variance of off-diagonal elements. Here,
because of Hermiticity, the matrix elements of $H^{R}$ should be
real and symmetric, and those of $H^{I}$ real antisymmetric
(therefore no diagonal elements in $H^{I}$) with respect to the
initial and final pair state of an interacting pair. Note that
such a modification is impossible for the single-$j$ model and
therefore is irrelevant for the explanation of the effect at least
for this particular case.

As shown in Ref. \cite{BFP}, the violation of $T$-invariance in
the form (\ref{4.4}) slightly {\sl enhances} the effect of
predominance of $f_{0}$ rather than reduces it. This might be
understood since the Gaussian ensemble with zero mean averages out
all odd powers of $H^{I}$, eliminating whatever the direct result
of its presence could be and leaving on average only a
renormalization of the real part. Although eq. (\ref{4.4}) keeps
intact the total variance of the random Hamiltonian, the higher
even moments are increased. Thus, if there was a trend (of
different origin) of pushing a state with $J=0$ down, this trend
would be amplified by the inclusion of $H^{I}$.

Nevertheless, the concept of $T$-invariance may be relevant for
the problem we are interested in. Indeed, any state with $J\neq 0$
appears as a multiplet of degenerate states $|JM\rangle$ with
various projections. For any given $M\neq 0$, the state
$|JM\rangle$ violates the $T$-invariance by choosing the sense of
precession of the angular momentum vector around the quantization
axis. This is nothing but a spontaneous symmetry breaking when the
symmetry of the ground state is lower than that of the
Hamiltonian. As always in such situations, the symmetry is
restored by the degeneracy with other states of the same
multiplet. A physical branch of the excitation spectrum that
restores the symmetry is {\sl rotation} that change the
orientation with no price in energy (Goldstone mode). In this
sense the state with $J=0$ is indeed singled out.

\subsection{Statistical widths?}

Many authors, starting with Ref. \cite{BFP}, explored the idea
that the statistical predominance of $J_{0}=0$ states is
associated with the shape of the level density $\rho(E;J)$ for a
given value of $J$. We have already mentioned that two-body
interactions in a finite Hilbert space produce the level density
close to Gaussian in each $J$-class around the centroid of all
strength functions (\ref{3.6}) $F_{k}(E)$ for the basis states
$|k\rangle$ of this class. Then the spectrum has a centroid
\begin{equation}
\bar{E}_{J}=\frac{1}{d_{J}}{\rm Tr}_{J}H\equiv \langle
H\rangle_{J},                                    \label{4.5}
\end{equation}
and the statistical width
\begin{equation}
\sigma^2_{J}=\langle (H-\langle H\rangle_{J})^{2}\rangle_{J}.
                                                   \label{4.6}
\end{equation}
The widths found as a result of statistical spectroscopy
\cite{Kota,Wong} for a given realization, have to be averaged over
the random ensemble.

It is natural to assume that the $J$-class with a larger
statistical width has a greater chance to contain the ground state
(that does not mean that the inverse statement is also correct).
The results of Ref. \cite{BFP} show that in the $sd$ shell model
for 4 and 6 particles the statistical widths for $J=0$ are by
approximately 10\% larger than for $J=2$ and for other low values
of $J$. Here one needs to notice that, even if this correlation of
$\sigma_{J}$ and $f_{J}$ would be universally correct, we actually
would simply reformulate the original question in another
language, namely what is the reason for the greater statistical
width of the $J=0$ class. But, moreover, the correlation is not
sufficiently strong and not universal \cite{Cov}.

\begin{figure}
\begin{center}
\includegraphics[width=8 cm]{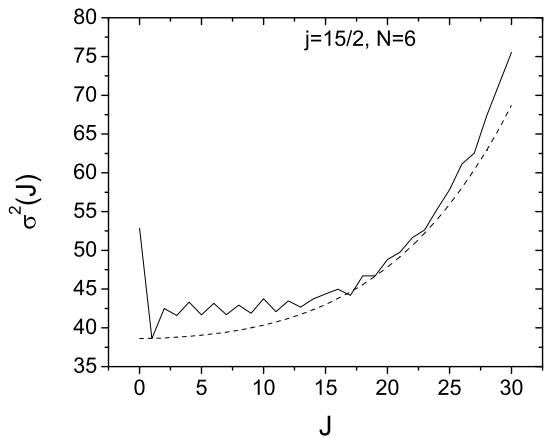}
\end{center}
\caption{\label{widthj15}
The average width $\sigma_{J}$ of the
level distribution in the $J$ class as a function of $J$ for 6
particles in a single level $j=15/2$, solid line; predictions of
the statistical formula, dashed line. }
\end{figure}

\begin{figure}
\begin{center}
\includegraphics[width=8 cm]{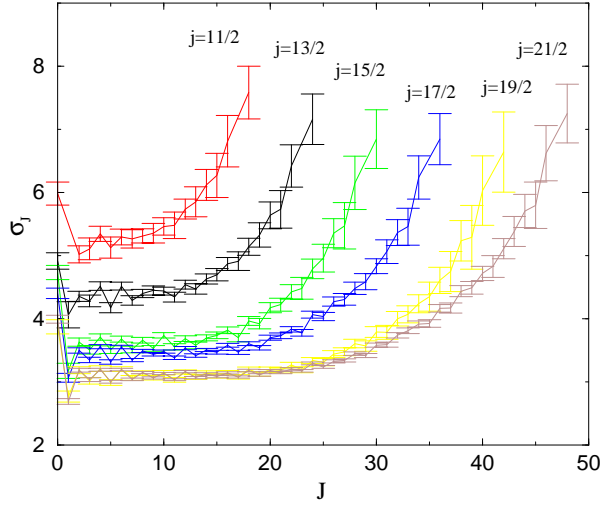}
\end{center}
\caption{\label{wids}
Widths $\sigma_{J}$ for various values of
the level spin $j$ in the single-$j$ model as a function of $J$
for 6 particles.}
\end{figure}

Figs. \ref{widthj15} and \ref{wids} illustrate the situation with
statistical widths in the single-$j$ models. Typically,
$\sigma_{0}$ is greater than the widths of competing low values of
$J$ but the widths $\sigma_{J}$ invariably grow higher than
$\sigma_{0}$ for many large values of $J$. However, only the
fraction of $J_{0}=J_{{\rm max}}$ is noticeably enhanced (still
never on the level more than 15\%) against multiplicity
expectations. One can show that the difference in the widths has
to be much greater, than it is in reality, in order to ensure, at
comparable multiplicities $d_{J}$, the observed predominance of
$J_{0}=0$ states. More subtle effects might be related
\cite{KapPJ} with deviations of higher moments of the statistical
distribution from Gaussian ones, especially for a system of spins
1/2.

The comparisons based on pure statistical characteristics of
individual $J$-classes are dangerous since they do not account for
correlations between the classes which are at the core of the
effect. If, for a given realization of the Hamiltonian,
$g_{J}(E)=\rho(E;J)/d_{J}$ is the level density for the $J$-class
normalized to unity, the probability of finding the levels of a
given class at energy higher than $E$ can be written as
\begin{equation}
\chi_{J}(E)= \left(\int_{E}^{\infty} dE'\,g_{J}(E')
\right)^{d_{J}}.                              \label{4.7}
\end{equation}
Then the probability of having the state $J$ below all other
states will be
\begin{equation}
f_{J}=\int_{-\infty}^{\infty}dE\,\left[-\frac{d}{dE}\chi_{J}(E)\right]
\prod_{J'\neq J}\chi_{J'}(E).                \label{4.8}
\end{equation}
The densities of different classes in this formula are strongly
correlated being determined by the same interaction. The task of
averaging this many-point correlation function over the ensemble
of random interactions is hardly solvable.

\section{Mesoscopic effects of geometry}

\subsection{General idea}

To find out how geometry of Hilbert space and of the random
angular momentum coupling can induce the observed effects, we can
first think of even simpler many-body  problems \cite{Mul,Yad}.
Take, for instance, a set of $N$ identically interacting spins,
\begin{equation}
H=\frac{1}{2}A\sum_{a\neq b}({\bf s}_{a}\cdot{\bf s}_{b}).
                                             \label{5.1}
\end{equation}
The energy spectrum of this system depends only on the total spin,
${\bf S}=\sum_{a}{\bf s}_{a}$,
\begin{equation}
E(S)=\frac{A}{2}[S(S+1)-Ns(s+1)],           \label{5.2}
\end{equation}
where $s$ is the single-particle spin. If in a random ensemble
the interaction strength
$A$ is symmetrically distributed with respect to zero we
see immediately that the ground state spin will be
either $0$ (antiferromagnetic ordering, $A>0$) or the maximum spin
$S=Ns$ (ferromagnetic ordering, $A<0$), both values appearing with
probability $f_{0}=f_{{\rm max}}=1/2$.

In this primitive example the answer is simple because the
spectrum of the system is pure rotational and all other quantum
numbers, except for the exact constant of motion, $S$, are not
differentiated by the interaction. We can expect that in all
cases, when the coupling of individual spins plays a role,
rotational modes (as we discussed above, they are Goldstone
excitations restoring the orientational invariance) with the most
ordered spin coupling schemes will bring in an enhanced
probability for the lowest and the highest value of the ground
state spin. The resulting probabilities are decided by the
statistical weights of regions in the parameter space with
positive and negative signs of the moment of inertia. The similar
situation takes place in the case of pairing that is in fact
rotation in gauge space: the energy of a state of seniority $s$ (a
number of unpaired fermions) in the degenerate model for $N$
particles is \cite{Talmi}
\begin{equation}
E(s)=\frac{G}{4}(N-s)(2\Omega-s-N+2),         \label{5.3}
\end{equation}
where $\Omega$ is the space capacity. Again, the ground state is
either fully paired one, $s=0$, or ``antipaired" one, $s=N$,
depending on the sign of the pairing constant $G$. Instead of the
rotation operator one has here {\sl quasispin} that is the
generator of $SU(2)$ algebra made of the pair transfer operators
$P,P^{\dagger}$ and particle number operator $N$.

The situation is very similar in the case of isovector pairing on
a single level. Here the problem can also be solved exactly with
the use of $R(5)$ group formed by six isovector pair creation and
annihilation operators, three components of isospin vector and
total particle number \cite{hecht65a,hecht65b,ginocchio65}. The
Hamiltonian
\begin{equation}
H_{{\rm i.p.}}=V_{0\,1} \, \sum_{\tau=0,\pm 1} P^\dagger_\tau \,
P_{\tau}\,,                                    \label{5.3a}
\end{equation}
has energy eigenvalues
\begin{equation}
E=\frac{V_{0\,1}}{\Omega} \left[
\frac{1}{4}(N-s)(2\Omega-N-s+6)+t(t+1)-T(T+1)\right]\,,
                                              \label{ISO:onel}
\end{equation}
here as before $s$ is a number of unpaired nucleons while $t$
denotes their total isospin.
The presence of the exactly conserved $SU(2)$ subgroup of isospin
makes this example particularly interesting. The correlations
between states with different isospin are represented by the term
$T(T+1)$ which indicates the presence of rotational $V_{0\,1}<0$
or antirotational $V_{0\,1}>0$ bands with 50-50\% probability. In
quadrupole boson models a similar situation occurs often \cite{BF}
with the presence of two ``rotational" Casimir operators,
three-dimensional, $\propto J(J+1)$, and five-dimensional,
$\propto v(v+3)$, where the boson seniority $v$ characterizes the
$O(5)$ group.

In a general case of complicated fermion dynamics, we can expect
that the geometric chaoticity will on average single out global
rotational modes, so that one can speak about an average
Hamiltonian that describes the relative positions of the {\sl
classes} of states with different exact quantum numbers, such as
total spin and isospin. The same logic works in the case of a
similar problem of the ground state spin in chaotic quantum dots
\cite{Folk}. In our case the effective Hamiltonian $\tilde{H}$ of
classes will take the form of the expansion in powers of the
scalar constant of motion ${\bf J}^{2}$,
\begin{equation}
\tilde{H}=H_{0}+H_{2}{\bf J}^{2}+H_{4}({\bf J}^{2})^{2} +\,...
                                          \label{5.4}
\end{equation}
The coefficients in this expansion are functions of the random
parameters in the original Hamiltonian. In the presence of
additional conserved quantities, as isospin, a similar expansion
in powers of ${\bf T}^{2}$ is to be added to (\ref{5.4}). The
problem is in (approximate) calculation of the coefficients in
this expansion and their averaging over random parameters. If, as
happens in reality, the quadratic term $H_{2}$ dominates, it
determines a rotational band with a random moment of inertia, and
the situation is analogous to that in eq. (\ref{5.2}). For
$J_{{\rm max}}$, the expansion (\ref{5.4}) may not work, and a
special treatment may be needed. This program was first
implemented in Refs. \cite{Mul,oddA,Yad}. A very similar
consideration independently and with different ideology was
carried out for interacting bosons \cite{BF}; we first comment on
the boson problem.

\subsection{Boson correlations}

Complicated fermion dynamics generate boson-like collective
excitations. Various types of phonons, magnons and plasmons are
just a few examples for macroscopic systems; shape vibrations and
giant resonances are well studied in mesoscopic physics  of nuclei
and atomic clusters. Because only few branches of the spectrum of
elementary excitations are collectivized, the {\sl bosonization}
of the many-fermion problem can lead to significant
simplifications of the formalism and a more transparent physical
picture. This is confirmed by many successes of the IBM
\cite{IBM}, where the boson model is postulated including certain
collective modes although their exact relation to the original
fermion interaction is not rigorously derived. The regular methods
of boson expansion of fermion operators were introduced in nuclear
physics long ago, \cite{BZ1} for a single-$j$ level and \cite{BZ2}
for a general level scheme, see the detailed review article
\cite{MK}. The application of the boson picture to our problem
seems promising, especially because numerous studies of the IBM
with random interactions \cite{BF,Bij,KZC,Kus,KSJ,Zhao2} found a
similar pattern of the predominance of $J=0$ ground states.

The IBM with two types of interacting bosons has \cite{IBM} the
parameter space sharply divided between the spheres of influence
of different symmetries. For example, this can be illustrated by
the peaks of the ICE \cite{cej} clearly marking the narrow
transitional regions between the symmetries. Using an ansatz of
the axially symmetric coherent intrinsic state generated by a
mixture of bosons, $sp$ for the vibron model and $sd$ for a
conventional nuclear IBM, as a trial function for the ground state
\cite{BF}, one can find the boundaries between the domains of
different symmetry. The coherent state corresponds to the
body-fixed frame with a certain orientation and undetermined spin;
parity is also violated in the vibron intrinsic state. Then one
needs to project out correct angular momentum states (it would be
better but more complicated to minimize the trial energy {\sl
after} this projection) and find the energy spectrum in the
space-fixed frame. A system of $p$-bosons is especially simple
since here quantum numbers $N$ and $J$ determine the state
uniquely (allowed total spins have the same parity as the boson
number and each spin appears once). A similar geometrical pattern
takes place with respect to isospin in odd-odd nuclei \cite{oddA},
where the same statistics is valid for quasideuteron pairs with
$L=0, \,t=1$; as a result some empirical regularities emerge with
random interactions.

As follows from Ref. \cite{BF}, the different intrinsic shapes of
the ground state, markedly separated in the space of trial
parameters, correspond to different coupling schemes of angular
momentum. Typically one gets, for an even boson number, the
condensate of scalar bosons with the only value $J=0$ possible,
condensate of deformed bosons with the rotational spectrum and the
probability divided between $J_{0}=0$ and $J_{0}=J_{{\rm max}}$
according to the sign of the moment of inertia, and the condensate
of multipole, dipole or quadrupole, quanta again with the same
alternative. Taking into account the corresponding areas in the
parameter space, one comes to a good agreement with ``empirical"
(numerical) data for the random ensemble. One can think of the
used procedure as of a variational method of constructing the
effective Hamiltonian $\tilde{H}$, eq. (\ref{5.4}). In the $sd$
model $\tilde{H}$ includes, apart from the rotational term
$H_{2}J(J+1)$, also an above mentioned term $H'_{2}v(v+3)$. This
picture does not account for the cases with $J_{0}$ or $v_{0}$
different from the edge values but the fraction of such cases is
low. The IBM is however much simpler than fermion systems because
the Bose-statistics creates condensates that in many cases
regularize the angular momentum coupling.

Going to the fermion system, we can try to use the boson expansion
method. The boson representation of pair operators should be a
reasonable approximation at least for a {\sl dilute} system
\cite{Mul,KZC,KapPJ} with particle number $N$ much smaller than
the space capacity $\Omega$. It follows from the commutator
(\ref{3.22}) (for simplicity we use the single-$j$ model) that
indeed under such conditions the pair operators $P$ and
$P^{\dagger}$ have quasiboson properties:
\begin{equation}
[P_{L'\Lambda'},P^{\dagger}_{L\Lambda}]=\delta_{LL'}
\delta_{\Lambda\Lambda'}+ \; {\rm terms\; of\; order}\;
\frac{N}{\Omega}.                                \label{5.5}
\end{equation}
Thus, we can introduce the ideal bosons $B_{L\Lambda}$ [$L$ even
from 0 to $L=2j-1=\Omega-2$] and find the boson expansion in
the (symbolically written) form
\begin{equation}
P_{L\Lambda}\approx B_{L\Lambda}+ [B^{\dagger}BB]_{L\Lambda}
+\,\dots.                                        \label{5.6}
\end{equation}

In the crudest boson approximation, the boson image of the
fermionic Hamiltonian is therefore the gas of $N_{B}=N/2$
noninteracting bosons with quantum numbers $L,\Lambda$ and
energies equal to $V_{L}$,
\begin{equation}
H_{B}^{(0)}=\sum_{L({\rm even})}V_{L}B^{\dagger}_{L}B_{L}.
                                                  \label{5.7}
\end{equation}
The ground state of $H_{B}^{(0)}$ is a condensate of all $N$
bosons in a mode with spin $L$ corresponding to the minimum
$V_{L}$. With random choice of $V_{L}$, this leads to a preference
for the ground state spin $J_{0}=0$ \cite{Mul}. Indeed, assume
that every $V_{L}$ has the same chance $1/k$ to be the smallest
one. If the value of $L$, $L({\rm min})$, corresponding to the
smallest $V_{L}$ equals zero, the total spin $J$ can be only zero
as well, the case analogous to the $s$-condensate in the IBM. All
other choices of $L({\rm min})$ create many degenerate states with
energy $E=N_{B}V_{L({\rm min})}$ and various values of total $J$
allowed for a given number of bosons with $L=L({\rm min})$,
including again $J=0$ on more or less equal footing. Summarizing
all cases, we should obtain a $J_{0}=0$ preference. The degeneracy
will be lifted by the interactions coming from higher boson
expansion terms.

The situation is illustrated by Fig. \ref{N4N5}d, where the
distribution of ground state spins is shown, solid line, for an
ensemble of 5 noninteracting bosons with random energies $V_{L}$
for $L=0,2,4,6,8$ and 10. The dotted line shows that the
statistical distribution of multiplicities is similar to that in
the Fermi-case. The predominance of $f_{0}$ is clearly seen being
however lower here than for fermions. Since the fraction $\sim1/k$
of the pure $s$-condensate falls off for larger systems, we
conclude that the simple bosonic effect does exist but cannot
explain the entire picture. A similar result is seen in Table
\ref{tab1} where the approximation of a 6-fermion system on a
$j=21/2$ level with 3 bosons is shown in column $(b)$.

\begin{figure}
\begin{center}
\includegraphics[width=11 cm]{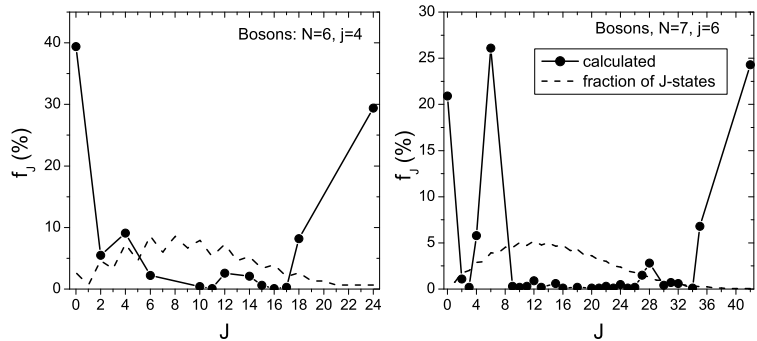}
\end{center}
\caption{ \label{N6N7}
The distribution of ground state spins for
bosons, $N=6,\,j=4$, left, and $N=7,\,j=6$ in the uniform ensemble
of interaction parameters.}
\end{figure}

\begin{figure}
\begin{center}
\includegraphics[width=14 cm]{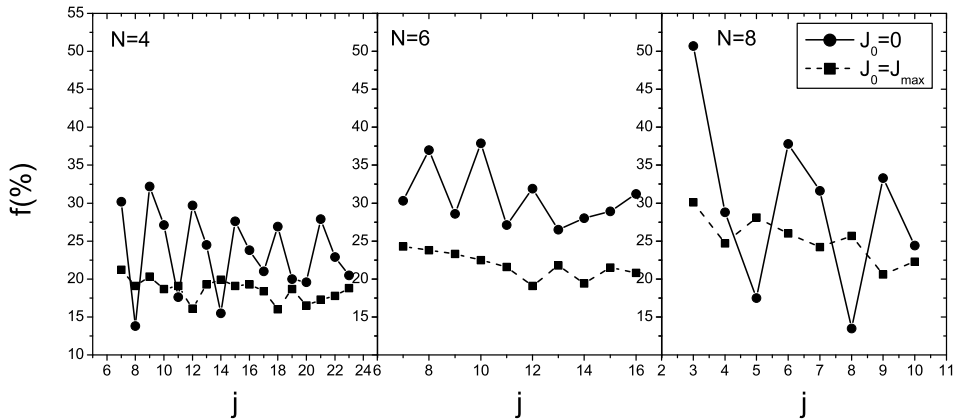}
\end{center}
\caption{ \label{pstatb} The systematics of probabilities $f_0$
for ground state spins $J_0=0$ (circles) and $J_0=J_{\rm max}$
(squares) in various systems of bosons; panels from left to right
show $N=4,6$, and 8 cases as a function of integer single-boson
spin $j$. }
\end{figure}

We can look also at the real interacting many-boson system
constructed in analogy to our single-$j$ fermionic cases, Figs.
\ref{N6N7} and \ref{pstatb}. The situation here is similar to the
IBM and shows that a considerable probability $f_{J}$ appears only
for $J=0$, $J=J_{max}=Nj$, and $J=j$, which can be explained in
the same way as in the IBM although the number of free parameters
grows with $j$. Despite similarities, the overall boson statistics
has some differences compared to the fermion case, see Figs.
\ref{pstat} and \ref{pstatb}, in particular the probability $f_0$
is generally lower, averaging around 25-35 \%. At the same time
the probability of an ``aligned'' ground state with $J_0=J_{\rm
max}$ is higher and for $N\ll 2j+1$ seem to stabilize at around
20-25\%.

\subsection{Sampling method and integrable systems}

Table 1, column (f), shows the predictions for ground state spins
in a system of $N=6,\,j=21/2$ obtained with the use of the
practical recipe suggested by Zhao, Arima and Yoshinaga
\cite{Zhao1}. The results here are produced by the consecutive
choice of the realizations of the random ensemble with only one
nonzero parameter $V_{L}<0$. In these cases, the absolute value of
this parameter is irrelevant. Thus, one performs the sampling of
the {\sl ``corners"} of the parameter space. For each such extreme
case, the diagonalization determines the ground state spin. The
final prediction for the fraction $f_{J}$ comes as a fraction of
the corners leading to $J_{0}=J$. Although this idea can be used
for fast numerical estimates, it does not shed light on the
physical reasons for the predominance of $J_{0}=0$ just adding
another interesting observation.

The closest analysis of the recipe suggested in ref. \cite{Zhao1}
shows that the results of this procedure are nor always right,
even qualitatively. A hard problem for such an empirical approach
emerges for the ensembles with different weights for different
matrix elements. The procedure does not indicate how the empirical
fractions obtained for the corners are to be modified for such
cases.

On the other hand, the corner evaluation indeed works in specific
cases of {\sl integrable} systems. Those are the cases when the
eigenstates of the Hamiltonian can be labeled by exact quantum
numbers as the consequence of rotational and some additional
symmetries. It is known, for example, that for the single level
$j=7/2$ all possible interactions preserve seniority, and the
eigenstates are uniquely characterized by spin and seniority.
Since the eigenfunctions do not depend on the interaction, the
energy eigenvalues are {\sl linear} functions of the interaction
constants \cite{Zhao}, and the coefficients of the linear form are
still determined by geometry of angular momentum coupling. The
search for the set of quantum numbers which provide the minimum of
such a function is reduced to a problem of linear programming. An
elegant geometric method of solving this problem was suggested by
Chau {\sl et al.} \cite{Chau}. It is clear that the knowledge of
corner energies is sufficient for this purpose.

The situation similar to that for fermions on $j=7/2$ orbit
appears in systems of $p,d$ or $f$ bosons with interaction
conserving the boson number. Again the appropriate quantum numbers
are total spin and boson seniority. The analogous case was
mentioned in the section on boson correlations for the systems
that were not pure in boson composition (for example, $s$ and $p$
or $s$ and $d$ bosons) but were considered with the aid of the
coherent variational function that introduced the condensate of a
specific boson mixture. The result was more complicated because
the proportions of mixture could depend on the interaction that
would make the problem nonlinear. Almost in all cases one again
sees the predominance of the states with extreme values of total
spin or/and seniority.

\section{Statistical approach}

\subsection{Effective Hamiltonian}

In order to come to an estimate of average ground state energy for
randomly interacting fermions we make a simple assumption that
there is a mean field generated by the interactions in each
realization of the ensemble and the field keeps axial symmetry so
that one can speak about mean occupation numbers $n_{m}$ for
fermions in an orbital with $j_{z}=m$ along the symmetry axis
(again we limit ourselves by a formally simpler case of a
single-$j$ level that still keeps the main features of the general
problem). As discussed in Section 2.4, the interaction leads to
equilibration so that the complicated states still can be
characterized by the single-particle occupation numbers. We do not
assume in advance thermal or other specific form of the
equilibrium distribution; instead we can derive it from the
simplest statistical arguments minimizing the ground state energy.
In fact, we just assume that the ground state is as chaotic as
excited states for the majority of realizations. This assumption
is in line with what we had seen in the analysis of the overlaps
of the actual ground state wave functions with fully paired
functions and with the realistic shell model.

In the mean field approximation, the expectation value of the
Hamiltonian (\ref{3.11}) in a statistical state described by the
occupation numbers $n_{m}$ can be written as
\begin{equation}
E(\{n_{m}\})=\frac{1}{2}\sum_{mm'}V_{mm'}\langle n_{m}n_{m'}
\rangle,                                        \label{6.1}
\end{equation}
where the amplitudes $V_{mm'}$ include the CGC,
\begin{equation}
V_{mm'}= 2\sum_{L\Lambda}V_{L}\left(C^{L\Lambda}_{jm\,jm'}
\right)^{2}.                               \label{6.2}
\end{equation}
Strictly speaking, eq. (\ref{6.1}) contains the correlated
occupation numbers taken for a given realizations and then
averaged. In the simplest approximation we substitute this by the
product $\langle n_{m}\rangle\langle n_{m'}\rangle$ of mean
occupation numbers; later the sign of averaging will be omitted.

The occupation numbers are subject to constraints due to the
conservation laws of the particle number $N$ and total angular
momentum projection $M$,
\begin{equation}
\sum_{m}n_{m}=N, \quad \sum_{m}mn_{m}=M.      \label{6.3}
\end{equation}
Considering fully aligned states we identify the projection $M$
with total spin $J$. This is similar to what is routinely done in
the nuclear cranking model when applied to the ``rotation" around
the symmetry axis, see for example \cite{Good1}. In this case the
angular momentum is explicitly built up by individual momenta of
the constituents, in keeping with the main idea of geometrical
chaoticity. Thus, we need to minimize the functional
\begin{equation}
\tilde{E}=\frac{1}{2}\sum_{mm'}V_{mm'}n_{m}n_{m'}-\mu\sum_{m}n_{m}
-\gamma\sum_{m}mn_{m},                         \label{6.4}
\end{equation}
where we added the constraints (\ref{6.3}) with Lagrange
multipliers of chemical potential, $\mu$, and cranking frequency
(or magnetic field), $\gamma$.

The extremum of the functional $\tilde{E}(\{n_{m}\})$ is at the
set $\{n_{m}\}$ that satisfies a system of linear algebraic
equations
\begin{equation}
\sum_{m'}V_{mm'}n_{m'}-\mu-\gamma m=0.          \label{6.5}
\end{equation}
By solving for $n_{m}(\mu,\gamma)$ and applying the constraints
(\ref{6.3}), we find the appropriate values of the Lagrange
multipliers $\mu(N,M)$ and $\gamma(N,M)$. The kernel (\ref{6.2})
of this inhomogeneous equation contains random parameters and
therefore in general can be inverted. The solution has a form of a
linear polynomial in $m$ (constant occupation for $M=J=0$ and
constant tilt for nonzero $J$). Finally, the value of energy for
this solution is
\begin{equation}
E(N,M)=\tilde{E}(\{n_{m}(\mu(N,M),\gamma(N,M))\}) +\mu(N,M)N
+\gamma(N,M)M.                                  \label{6.6}
\end{equation}
Moreover, since eq. (\ref{6.5}) is still satisfied by this
specific choice of $\mu$ and $\gamma$, it leads to
\begin{equation}
\sum_{mm'}V_{mm'}n_{m}(N,M)n_{m'}(N,M)=\mu(N,M)N+\gamma(N,M)M,
                                              \label{6.7}
\end{equation}
that determines the energy at the extremum in the simple form
\begin{equation}
E(N,M)=\frac{1}{2}[\mu(N,M)N+\gamma(N,M)M]        \label{6.8}
\end{equation}
that does not require an explicit expression for the occupation
numbers. Of course, we just applied a standard procedure used in
thermodynamics for the Legendre transformation between different
potentials.

The value of the chemical potential can be found in a general way
without actually solving the set of equations (\ref{6.5}). Summing
those equations for all $m$ and taking into account that
\begin{equation}
\sum_{m}1=2j+1\equiv \Omega, \quad \sum_{m}m={\rm Tr}\,j_{z}=0,
                                                 \label{6.9}
\end{equation}
we obtain
\begin{equation}
\mu=\frac{1}{\Omega}\sum_{mm'}V_{mm'}n_{m'}=\frac{2N}{\Omega^{2}}
\sum_{L}(2L+1)V_{L},                             \label{6.10}
\end{equation}
where we used the normalization of the CGC,
\begin{equation}
\sum_{m\Lambda}\left(C^{L\Lambda}_{jm\,jm'}\right)^{2}=\frac{2L+1}
{\Omega}.                                       \label{6.11}
\end{equation}

In a similar way we can calculate the cranking parameter $\gamma$.
Multiplying eqs. (\ref{6.5}) by $m$ and summing over $m$, we come
to
\begin{equation}
\gamma=\frac{\sum_{mm'}mV_{mm'}n_{m'}}{\sum_{m}m^{2}}=\sum_{L}V_{L}
\sum_{m'}n_{m'}y_{L}(m'),                         \label{6.12}
\end{equation}
where the geometric factor can be calculated as
\begin{equation}
y_{L}(m')\equiv 2\sum_{m\Lambda}m\left(C^{L\Lambda}_{jm\,jm'}
\right)^{2}=\alpha_{L}m',                        \label{6.13}
\end{equation}
\begin{equation}
\alpha_{L}=\frac{{\bf L}^{2}-2{\bf j}^{2}}{{\bf L}^{2}}\,
\frac{\sum_\Lambda\Lambda^{2}}{\sum_{m}m^{2}}.     \label{6.14}
\end{equation}
Since
\begin{equation}
\sum_{m}m^{2}=\Omega\,\frac{{\bf j}^{2}}{3}, \quad \sum_\Lambda
\Lambda^{2}=(2L+1)\,\frac{{\bf L}^{2}}{3},       \label{6.15}
\end{equation}
the final result reads
\begin{equation}
\gamma=\frac{3}{\Omega^{2}{\bf j}^{4}}\sum_{L}(2L+1)({\bf L}^{2}
-2{\bf j}^{2})\,M.                               \label{6.16}
\end{equation}
The geometric meaning of the combination in eq. (\ref{6.16}) that
determines the sign of the contribution of pairs with spin $L$ can
be easily understood. By definition (6.4), for $M>0$, the energy
(\ref{6.8}) in the laboratory system goes to minimum for negative
$\gamma$. The inequality ${\bf L}^{2}>2{\bf j}^{2}$ means that the
constituents of the pair are aligned, and this contributes to the
reduction of energy if there is {\sl attraction}, $V_{L}<0$, for
this component of the interaction. Vice versa, there should be
{\sl repulsion}, $V_{L}>0$, for antialigned pairs, ${\bf L}^{2}>
2{\bf j}^{2}$.

Identifying $M$ with the magnitude $J$, we come to the effective
Hamiltonian (\ref{5.4}). In the statistical approximation,
$\tilde{H}$ consists of only two terms, $H_{0}$ and $H_{2}$. The
rotational term $H_{2}$ is in this approximation {\sl linear} in
random parameters. Therefore the crude prediction is that the
probability of having the ground state spin $J_{0}=0$ is 1/2.
Although the effective moment of inertia given by the inverse
coefficient in front of $M$ in the parameter $\gamma$, eq.
(\ref{6.16}), does not depend on particle number, the statistical
approach has to work better for a larger $N$. This is indeed seen
in Fig. \ref{pstat} where we juxtapose the results for $f_{J}$ in
the systems of $N=4,6$ and 8.

One can also note \cite{Yad} that the resulting prediction for the
two items in energy, eq. (\ref{6.8}), picks up the monopole,
$K=0$, and dipole, $K=1$, terms in the Hamiltonian written in the
multipole-multipole form (\ref{3.15}) with an additional factor of
two and the interaction parameters given by the transformation to
the particle-hole channel (\ref{3.17}); the $L$-dependence in the
effective moment inertia (\ref{6.16}) comes from the $6j$-symbol
in eq. (3.17) for $\tilde{V}_{1}$. The extra factor of two
originates from two possible recouplings of single particle
operators in the two-body Hamiltonian on the way to the
statistical approximation (\ref{6.1}). The $K=0$ and $K=1$
multipole interactions are not independent from higher multipoles
[eq. \ref{3.19}]. In the statistical limit the monopole term is
produced in half by $\tilde{V}_0$ in eq. \ref{3.15}, while the
other half comes from all higher $\tilde{V}_K$ combined via eq.
\ref{3.21}, see also discussions and examples in \cite{Volya}.

\begin{figure}
\begin{center}
\includegraphics[width=14 cm]{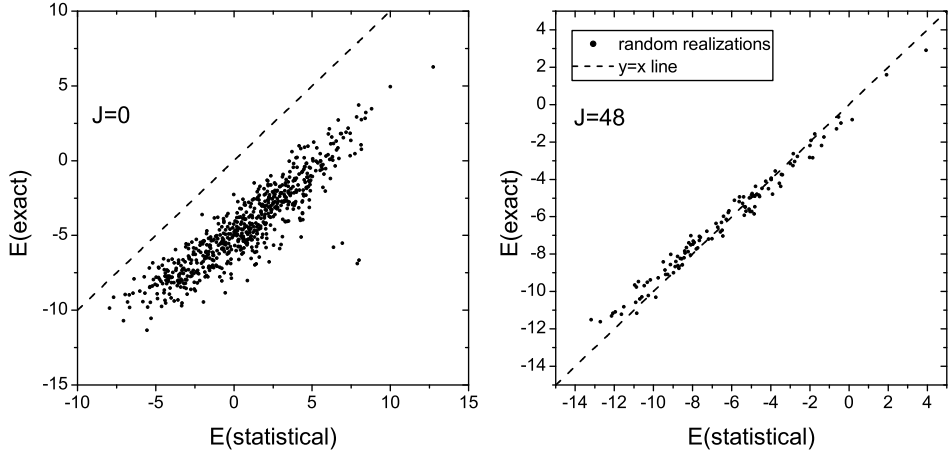}
\end{center}
\caption{\label{esj}
For a system of $N=6,\,j=21/2$, individual
realizations with $J_0=0$, left panel, and $J_0=J_{{\rm Max}}$,
right panel, are presented as points with coordinate $x$ equal to
the value of energy from the statistical prediction and
$y$-coordinate corresponding to the exact energy from
diagonalization.}
\end{figure}

\begin{figure}
\begin{center}
\includegraphics[width=7 cm]{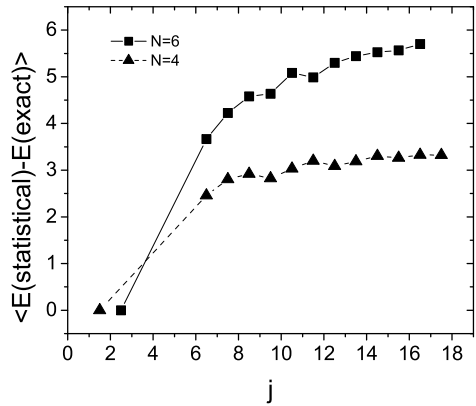}
\end{center}
\caption{\label{shift} Average difference in energy between exact
diagonalization and statistical prediction for systems of $6$
particles as a function of $j$.}
\end{figure}

The quality of the statistical description can be inferred from
correlations of actual ground state energy in a given copy of the
ensemble with the statistical value, Fig. \ref{esj}. For the
maximum possible momentum the statistical formula works nearly
perfect, for the $J=0$ states the overall correlation is very
good. However, exact energy is shifted down by a constant, which
indicates correlations beyond the statistical description. This
energy shift is shown in Fig. \ref{shift} as a function of the
size of the space $j$ for a six particle system.

\subsection{Occupation numbers}

The equivalence of the ``monopole + dipole'' truncation to the
results of the lowest statistical approximation means that higher
multipole interactions, being not associated with any conserved
quantities, on average do not influence the equilibrium occupation
numbers. This cannot be always true. Higher multipole interactions
responsible for real deformation of the mean field lift the
remaining degeneracies, compare the bosonic case. However, as $N$
and $\Omega$ grow, many competing multipoles tend to cancel each
other. Therefore we expect the validity of the statistical
approximation to improve as well. Essentially the same result can
be derived by the direct calculation of the cranking model moment
of inertia.

\begin{figure}
\begin{center}
\includegraphics[width=7 cm]{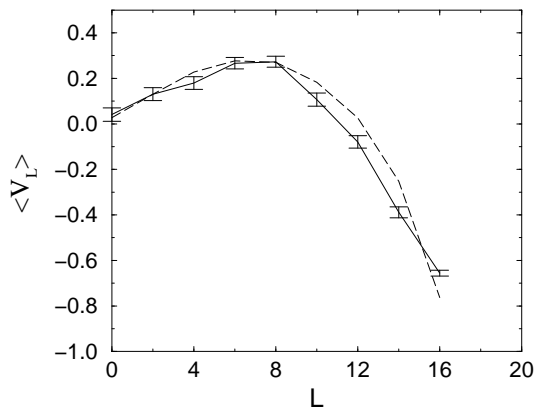}
\end{center}
\caption{\label{N17}
Mean values $\langle V_{L}\rangle$ of
interaction parameters for the cases with $J_{0}=J_{{\rm max}}$ in
the $N=6,\,j=17/2$ system; numerical simulations, solid line, and
statistical predictions, dashed line.}
\end{figure}

At the same time, the result above does not mean that the
probability of $J_{0}=J_{{\rm max}}$ also goes to 50\%. The
multiplicity of the class with $J=J_{{\rm max}}$ is very low - in
the single-$j$ model there exists only one such state with the
unique full alignment of all available particles. Although this
state by itself can be described well with the statistical
approach, we cannot reliably compare the energy of this state with
energies of states with other spins split due to the higher
multipole interactions. The fact that the statistical description
works for $J=J_{{\rm max}}$ is seen from Fig. \ref{esj} and Fig.
\ref{N17}, where the average values of the parameters, $\langle
V_{L}\rangle$, obtained from the ensemble copies that resulted in
$J_{0}= J_{{\rm max}}$, are compared to the statistical
predictions.

In Ref. \cite{Mul} a stronger statistical assumption was made. The
occupation numbers $n_{m}$ were modeled by those in a Fermi gas
with high temperature when one can neglect dynamical splitting of
single-particle orbitals $|jm)$, and the occupation numbers are
determined solely by the constraints (\ref{6.3}). In this case one
can take
\begin{equation}
n_{m}=\frac{1}{1+\exp(-\alpha-\beta m)},      \label{6.17}
\end{equation}
which ensures the statistical demands, $0\leq n_{m}\leq 1$, and
includes, as in eq. (\ref{6.4}), two parameters associated with
the conservation laws (\ref{6.3}). For any state with $J=M=0$, all
orbitals $|jm)$ must have the same occupancy,
\begin{equation}
n_{m}=\langle 00|a^{\dagger}_{m}a_{m}|00\rangle=
\frac{N}{\Omega}\equiv \bar{n}.                  \label{6.18}
\end{equation}
$\beta=0$, and the corresponding constant $\alpha_{0}$ is related
to the particle number as
\begin{equation}
\alpha_{0}=\ln\,\frac{\bar{n}}{1-\bar{n}},    \label{6.19}
\end{equation}
vanishing for the half-filled shell when $\bar{n}=1/2$. We usually
try to avoid such systems which are exceptional, see for example
\cite{Zhao1}, because of particle-hole symmetry that eliminates
even-$K$ multipole moments.

\begin{figure}
\begin{center}
\includegraphics[width=14 cm]{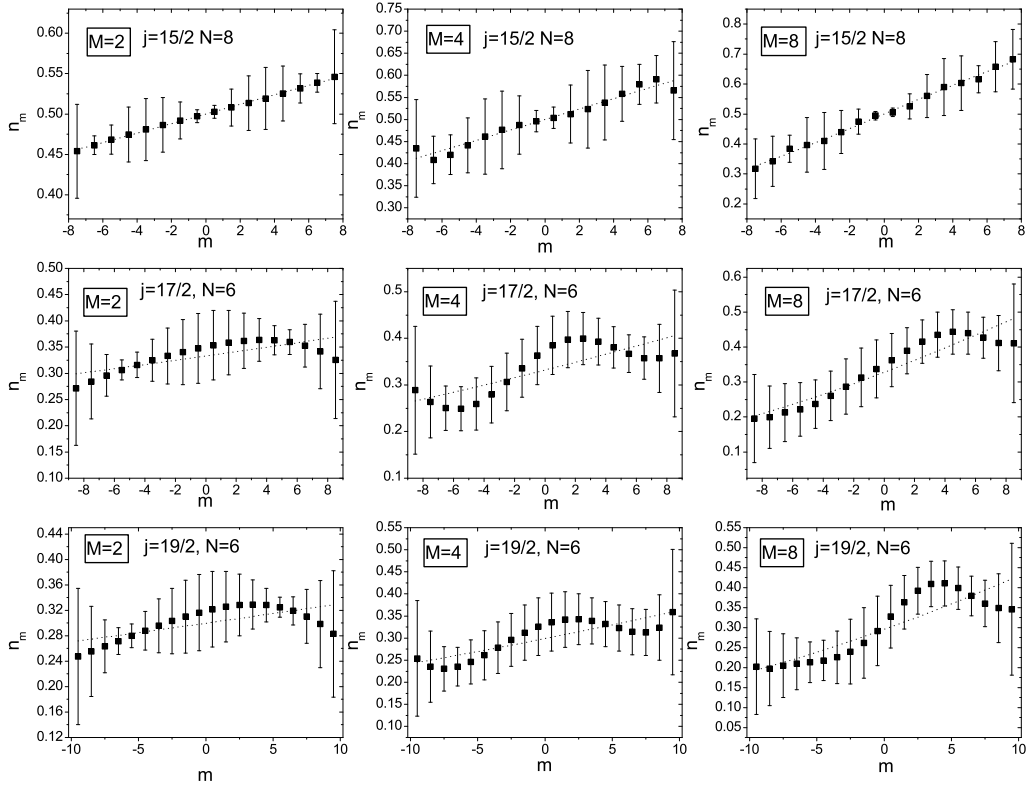}
\end{center}
\caption{\label{nmall} Average occupation numbers of yrast states
of a given spin $J$ and projection $J_{z}=M=J$ for the
$N=8,\,j=15/2$ system, top row; for $N=6,\,j=17/2$, middle row;
and $N=6,\,j=19/2$, lowest row. Numerical results are shown with
squares with error bars indicating rms deviations; statistical
predictions are indicated with dotted lines.}
\end{figure}


For a nonzero total spin projection $M$, the parameter $\beta$,
similar to the cranking frequency $\gamma$, creates a tilt of the
function $n_{m}$. As shown in \cite{Mul}, the main effect is
caught by the expansion of the distribution function (\ref{6.17})
in powers of $\beta$. Due to time-reversal invariance, the
parameter $\alpha$ acquires a correction in the second order, and
the results are
\begin{equation}
\beta=\frac{M}{\bar{n}(1-\bar{n})}\sum m^{2},     \label{6.20}
\end{equation}
\begin{equation}
\alpha=\alpha_{0}+\alpha_{2}, \quad \alpha_{2}=\left(\bar{n}-
\frac{1}{2}\right)\beta^{2}\frac{\sum m^{2}}{\Omega}. \label{6.21_1}
\end{equation}
In accordance with chaotic angular momentum coupling, a nonzero
spin $M$ is created by the fluctuations of occupancies, $\propto
\bar{n}(1-\bar{n})$, as in standard theory of level density of a
Fermi-gas \cite{Bethe,Eric}. The occupancies now can be written as
\begin{equation}
n_{m}=\bar{n}+\frac{mM}{\Omega\langle m^{2}\rangle}-\frac{\bar{n}
-1/2}{\bar{n}(1-\bar{n})}\,\frac{M^{2}}{\Omega^{2}\langle
m^{2}\rangle^{2}}\,(m^{2}-\langle m^{2}\rangle). \label{6.21}
\end{equation}

This approach works for bosons in the same way. In the case of $N$
bosons on a single $j$ level (integer $j$), the occupation numbers
in the same approximation are given by
\begin{equation}
n_{m}=\frac{1}{\exp(-\alpha-\beta m)},           \label{6.21a}
\end{equation}
and the parameters can be found as
\begin{equation}
\alpha=\alpha_{0}+\alpha_{2}, \quad \alpha_{0}=\ln\,\frac{\bar{n}}
{\bar{n}+1}, \quad \bar{n}=\frac{N}{\Omega},      \label{6.21b}
\end{equation}
\begin{equation}
\beta=\frac{M}{\bar{n}(1+\bar{n})}\sum m^{2},     \label{6.21c}
\end{equation}
\begin{equation}
\alpha_{2}=-\left(\bar{n}+\frac{1}{2}\right)\beta^{2}\frac{\sum
m^{2}}{\Omega}.                                    \label{6.21d}
\end{equation}
The final result is analogous to (\ref{6.21}):
\begin{equation}
n_{m}=\bar{n}+\frac{mM}{\Omega\langle m^{2}\rangle}+\frac{\bar{n}
+1/2}{\bar{n}(1+\bar{n})}\,\frac{M^{2}}{\Omega^{2}\langle
m^{2}\rangle^{2}}\,(m^{2}-\langle m^{2}\rangle), \label{6.21e}
\end{equation}
which predicts the change of curvature compared to the Fermi
expression (\ref{6.21}). Note that the real condensate of bosons
at a single $m$ value is impossible being in contradiction to the
angular momentum requirement.

The expansion (\ref{6.21}) is rapidly converging because of the
powers of the ``volume" $\Omega$ in the denominator. Using this
for evaluating energies (\ref{6.1}) of the states along the yrast
line \cite{Mul}, one again comes to the effective Hamiltonian in
the form (\ref{5.4}), where the scalar and quadratic terms
coincide with those found in the variational approach. The second
order correction in (\ref{6.21}) adds the quartic term
$H_{4}\propto ({\bf J}^{2})^{2}$. This contribution, which is
small at not very high $J$, was taken in \cite{Mul} to account for
the difference between the observed fraction $f_{0}$ and its
limiting statistical value of 1/2. Such corrections can be
obtained with an improved variational ansatz of the previous
subsection. As seen from Fig. \ref{nmall}, the actual occupation
numbers for given $J=M$, averaged over the ensemble of yrast
states with given $J$, indeed quite well follow the linear
$m$-dependence. There are deviations from the simplest statistical
and variational predictions. The last term in (\ref{6.21})
includes kinematic correlations due to the Fermi-statistics, eq.
\ref{6.17}, however it does not describe fully dynamical effects.
The case of a half-occupied system is particularly interesting:
here the fluctuations are significantly suppressed, see first row
in Fig. \ref{nmall}; accordingly, the term proportional to
$(m^{2}-\langle m^{2}\rangle)$ in eq. (\ref{6.21}) disappears.
However, a regular oscillatory behavior of the occupancies around
the mean statistical behavior survives.

\subsection{Multipole collectivity}

We have seen earlier that the structure of the ground state is far
from that of the paired condensate. The question if the set of
random interactions generates, along with the ground states of
zero spin, some collective structure of the excitation spectrum
was put forward already in the original paper \cite{JBD}. To find
out the answer, the authors looked at the saturation of
transitions from the ground state of $J_{0}=0$ to the first
excited state of $J=2$ for a particle-hole operators of quadrupole
type. They observed that it is possible to construct the
quadrupole operator that maximally connects two states, and the
resulting transition accumulates more than 50\% of the
corresponding sum rule, in similarity to the well known
collectivity of the first $2^{+}$ states in non-magic even-even
nuclei. In essence, this emphasizes a particle-hole nature of the
transition with the operator adjusted to each copy of the random
ensemble.

Actual quadrupole collectivity in nuclei is usually considered
\cite{BM} as a result of coherent interactions in the
particle-hole channel with $K=2$. The background created by
pairing is important since the low-lying phonon collective
excitation should be located within the energy gap due to pairing
\cite{Bel}. In a normal Fermi gas low-lying modes have nearly pure
single-particle character. In contrast to that, in superfluid
systems the presence of the gap stabilizes collective modes as
coherent superpositions of two-quasiparticle excitations. At
sufficiently strong collectivity, the mode found in the random
phase approximation (RPA) becomes unstable, and then effects of
anharmonicity lead to static deformation. In deformed nuclei, the
low-lying quadrupole modes give rise to rotations and new
vibrations around the deformed equilibrium point.

A comparison of ``normal" quadrupole collectivity with data from
the random interaction ensemble shows \cite{Hor} that collective
effects are strongly suppressed. One needs to "displace" the
ensemble including explicitly a coherent attractive part in order
to reproduce the collectivity \cite{vela,velah}, as was
illustrated long ago by Cortes, Haq and Zuker \cite{cortes}. The
fractional collectivity suggested in Ref. \cite{JBD} was
calculated for 8 particles in the $sd$-shell model for $^{24}$Mg
using the realistic interaction \cite{BW} and various random
ensembles mentioned above, Sect. 4.2. The degree of collectivity
was defined as
\begin{equation}
f.c.=\frac{B({\rm E2}; 0_{1}\rightarrow 2_{1})}{\sum_{n} B({\rm
E2};0_{1}\rightarrow 2_{n})},           \label{6.22}
\end{equation}
where the reduced probability $B({\rm E2})$ of the quadrupole
transition was determined with the {\sl fixed} quadrupole operator
rather than with different operators maximized for each set of
random parameters. This quantity is significantly smaller than
found in Ref. \cite{JBD} for adjusted operators.

\begin{figure}
\begin{center}
\includegraphics[width=7 cm]{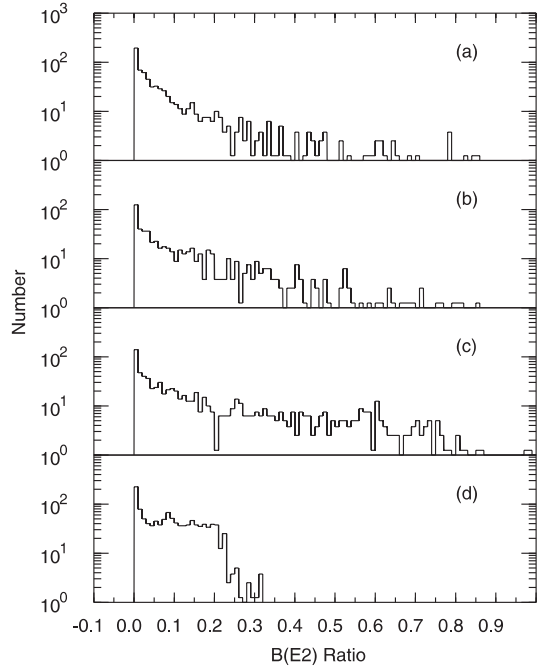}
\end{center}
\caption{ \label{prlfig2}Distribution of
probabilities $B({\rm E2}; 0_{1}\rightarrow
2_{1})$ in random ensembles $a-d$, Sect. 4.2, in units of the
probability in realistic shell model for $^{24}$Mg \cite{BW}.}
\end{figure}

Fig. \ref{prlfig2} shows calculated distributions of $B({\rm E2})$
for different ensembles (in units of the transition probability of
69.5 $e^{2}{\rm fm}^{4}$ found in the $sd$-shell model \cite{BW})
for $^{24}$Mg. Typically, the $B({\rm E2})$ values from random
interactions are by more than an order of magnitude weaker than in
realistic calculations. Even the maximum $B({\rm E2})$ values out
of 1000 samples for all four models are smaller than the realistic
value, although is few copies they come close. The distribution of
the $B({\rm E2})$ values for models (a) and (b) is close to the
Porter-Thomas (\ref{2.6}) as expected for matrix elements of a
simple operator between two chaotic states \cite{Brody,SF,ann};
the first excited state is even less regular than the ground
state. The model (c) with realistic pairing generates a hint of
collectivity. This agrees with what we have said above concerning
the role of pairing correlations supporting the multipole
collectivity. The sharp cutoff at small values of $B({\rm E2})$ in
model (d) happens close to the value that can be obtained for the
pure $(d_{5/2})_{p}^{4}(d_{5/2})_{n}^{4}$ configuration since here
the multipole-multipole correlations generated by the higher-$L$
components of interaction are absent.

We have seen, Fig. \ref{rpairing}, that, in spite of mainly
chaotic nature of eigenstates for random interactions, there
exists a slight excess of cases with a significant overlap between
the ground state wave function and the fully paired state. This
excess is particularly noticeable in cases when the sequence of
the lowest states is $J_{0}=0,\,J_{1}=2$ (such cases appear also
with a higher probability than could be expected from the
statistical multiplicity). This means that there is a probability,
exceeding the expectation of pure random models, that random
interactions indeed create collective effects. To illustrate this
point, we consider the dynamical quantity, which we call {\sl
Alaga ratio}, that can distinguish between different collective
structures,
\begin{equation}
A=\frac{{Q}^2}{B(\rm E2)}.                    \label{6.22a}
\end{equation}
Here the numerator is the expectation value squared of the
quadrupole moment of the first excited state $2_{1}$,
\begin{equation}
{Q}=\langle J M=J|{M}_{2 0}|J\, M=J\rangle\,  \label{6.23}
\end{equation}
and the $B({\rm E2})$ transition strength in the denominator is
defined as
\begin{equation}
B{\rm (E2)}=\sum_{M_f \kappa}\,|\langle J M_f|{M}_{2 \kappa}|J\,
M_i \rangle |^2\,.                             \label{6.24}
\end{equation}
Thus, limiting ourselves by the sequences  $J_{0}=0,\,J_{1}=2$, we
are looking at the ratio of diagonal to off-diagonal matrix
elements of the quadrupole operator.

\begin{figure}
\begin{center}
\includegraphics[width=14 cm]{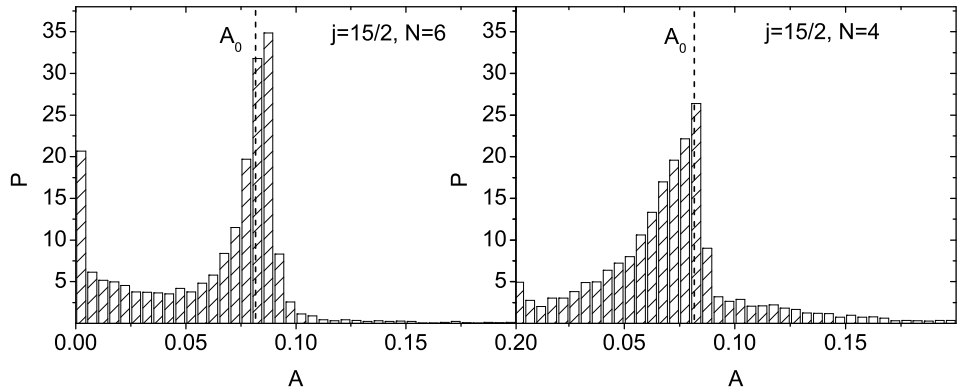}
\end{center}
\caption{ \label{lalaga} The Alaga ratio for 6 and 4 particles in
a single $j=15/2$ level; the histogram includes the states with
the sequence $J_{0}=0, J_{1}=2$ of lowest spins.}
\end{figure}

If random coupling of individual spins results in average
spherical shape, and the ground and the first excited state are of
similar structure, the Alaga ratio (\ref{6.22a}) should be small.
On the other hand, if this random coupling creates a more or less
rigid structure, one can expect the fulfillment of Alaga intensity
rules \cite{BM},
\begin{equation}
A=\frac{4}{49}\equiv A_{0}\,.                   \label{6.25}
\end{equation}
For the two lowest states without any genetic interconnection, the
Alaga ratio can take any value. In the case of single $j=15/2$
model, the sequences of interest appear in 6.7\% for $N=4$ and in
9.2\% cases for $N=6$. The distribution of the Alaga ratio for
these systems, Fig. \ref{lalaga}, reveals two peaks at $A=0$ and
$A=A_{0}$. The peaks are pronounced stronger at a larger particle
number. The idea that such structure can arise from random
interactions is not that surprising. It was proven long ago
\cite{Rock} that the effects of interactions in large
non-superfluid rotating Fermi-systems cancel leaving the
rigid-body moment of inertia. In agreement with this, the
statistical dependence of the level density, based on the
geometrical chaoticity, also gives the same value of the moment of
inertia, see \cite{Eric} and eq. (\ref{3.5}). The Alaga ratio
seems to be a more sensitive signature of rotational behavior than
the standard ratio of energies involving the next $J=4$ state.

\subsection{Multipole dynamics}

The multipole dynamics can be studied analytically starting with
the operator equations of motion for the multipole operators
$M_{K\kappa}$ and using the Hamiltonian in a suitable form
(\ref{3.15}) and commutation relations (\ref{3.24}). The exact
equations of motion are
\[[M_{K\kappa},H]=-g_{K}\sum_{K'S}g_{K'}g_{S}\tilde{V}_{K'}
[1-(-)^{K+K'+S}]\left\{\begin{array}{ccc}
                            K & K' & S\\
                            j & j & j\end{array}\right\}\]
\begin{equation}
\times\sum_{\kappa'\sigma}(-)^{\sigma}\left(\begin{array}{ccc}
                            K & K' & S\\
\kappa & \kappa' & -\sigma\end{array}\right)\frac{1}{2}
[M_{S\sigma},M^{\dagger}_{K'\kappa'}]_{+},      \label{6.26}
\end{equation}
where $[...\,,\,...]_{+}$ denotes an anticommutator, and
$g_{K}=\sqrt{2K+1}$. Note that zero values $K'$ and $S$ do not
contribute to these equations, and $K=0$ or 1 give trivial zero
results because of the conservation laws.

We take in eqs. (\ref{6.26}) the matrix element between the ground
state $|0\rangle$, assumed to have zero spin, and a hypothetical
collective state $|K\kappa\rangle$. After separating the dependence on
magnetic quantum numbers by the Wigner-Eckart theorem,
\begin{equation}
\langle S\sigma|M_{K'\kappa'}|K\kappa\rangle=(-)^{\sigma+\kappa'}
\left(\begin{array}{ccc} S & K' & K \\
-\sigma & -\kappa' & \kappa\end{array}\right)M^{K'}_{SK},
                                           \label{6.27}
\end{equation}
we come to the set of nonlinear equations for the matrix elements
\begin{equation}
M_{K}\equiv \langle 0|M_{K\kappa}|K\kappa\rangle\equiv (-)^{K}
\frac{1}{g_{K}}\,M^{K}_{0K},                  \label{6.28}
\end{equation}
that contain the excitation energy $E_{K}$ of the collective
state,
\begin{equation}
E_{K}g_{K}M_{K}=\sum_{K'S}g_{K'}g_{S}
(\tilde{V}_{K'}-\tilde{V}_{S})\frac{1-(-)^{K+K'+S}}{2} \left\{
\begin{array}{ccc}
K & K' & S \\
j & j& j \end{array}\right\}M_{K'}M^{S}_{K'K}. \label{6.29}
\end{equation}
Those equations are still exact if the sum runs over all allowed
intermediate states of spin $K'$.

In the spirit of the RPA we make here truncation leaving only the
most coherent contributions with $K'=K$ when the multipole $K$
does not share its angular momentum with other excitations.
Thereby the equation gets linearized, and we obtain a closed
expression for the excitation energy
\begin{equation}
E_{K}=\sum_{S({\rm odd})}g_{S} (\tilde{V}_{K}-\tilde{V}_{S})
\left\{\begin{array}{ccc}
K & K & S \\
j & j& j \end{array}\right\}M^{S}_{KK}.      \label{6.30}
\end{equation}
The result (\ref{6.30}) depends on expectation values $M^{S}_{KK}$
of odd-spin multipoles; the whole dynamics is concentrated in the
differences $\tilde{V}_{K}-\tilde{V}_{S}$ of multipole coupling
constants. The diagonal matrix elements of {\sl odd} multipoles of
rank $S$ can be constructed as those of irreducible tensors made
of the $S$ components of angular momentum ${\bf J}$, another
manifestation of fractional parentage or quasi-random geometrical
coupling.

Using again the commutators (\ref{3.24}) for odd $K'$ and applying
the same RPA-like approximation, we come to
\begin{equation}
M^{S}_{KK}=2g_{K}^{2}g_{S}\left\{\begin{array}{ccc} K & S &K\\
j & j & j\end{array}\right\}, \quad S =\,{\rm odd}. \label{6.31}
\end{equation}
In the semiclassical limit of large $j$ the $6j$-symbol in eq.
(\ref{6.31}) is proportional to the Legendre polynomial
$P_{S}(\cos\theta)$, where $\theta$ is the angle between ${\bf j}$
and ${\bf K}$. This describes the reorientation of the occupation
numbers in the process of collective excitation by the multipole
$K$. In this approximation, the collective excitation energy is
given by
\begin{equation}
E_{K}=2g_{K}\sum_{S({\rm odd})}g_{S}^{2}
(\tilde{V}_{K}-\tilde{V}_{S}) \left\{\begin{array}{ccc}
K & K & S \\
j & j& j \end{array}\right\}^{2}.      \label{6.32}
\end{equation}
The earlier discussed statistical limit, $E_{K}\approx
\tilde{V}_{1}K(K+1)$, is given by the contribution of the $S=1$
term. The other terms determine the collectivity corrections.

\begin{figure}
\begin{center}
\includegraphics[width=7 cm]{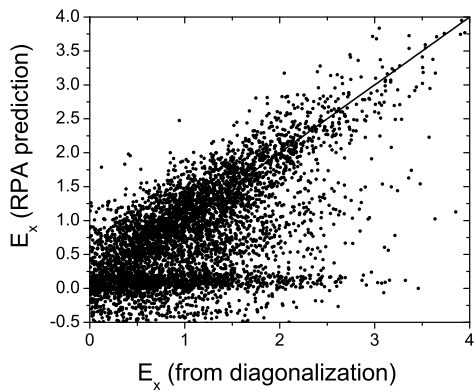}
\end{center}
\caption{\label{nrj15n6} Transition energies between the ground
state $J_{0}=0$ and the yrast state $J=K$ for a system
$N=6\,j=15/2$ in a random ensemble compared to the RPA
approximation, eq. (\ref{6.32}).}
\end{figure}

Fig. 15 shows the correspondence between the excitation energies
found in many runs with random interactions for the $N=6,\,j=15/2$
system and the RPA prediction (\ref{6.32}). The strong correlation
observed here confirms that, even for random interactions, the
geometry of space and spin coupling forms coherent combinations
that can be assessed with the regular methods of many-body theory.
Also a horizontal line of point accumulation is seen that
corresponds to the situation when the low-lying states are nearly
degenerate. It would be interesting to study what part of the
random parameter space corresponds to such absence of collective
correlations.

\section{Conclusion}

The idea to study the physics of a finite mesoscopic system with
random interactions turned out to be very fruitful. Many
theoreticians responded to the challenge; many new things were
learned even for the simplest systems of few identical particles;
many questions are not answered yet. After four years of extensive
studies the main result can be formulated as following: standard
textbook ideas of the factors that form the low-lying structure of
a closed self-sustaining mesoscopic systems are insufficient. The
quantum numbers of the ground states and some regularities of
spectra emerge not necessarily due to the corresponding coherent
parts of the interparticle interaction. Up to a large extent,
these characteristics are predetermined by the conservation laws
and geometry of available single-particle space.

The angular moment $J_{0}=0$ of the ground state in even-even
nuclei appears with a probability of the order 50\% with any
randomly taken rotationally invariant residual interaction. The
underlying mechanism may be related to chaotic coupling of
individual spins that creates an average yrast line described by
the effective Hamiltonian with mainly quadratic dependence on
total angular momentum. The time-reversible $J=0$ state turns out
to have an exceptional ability of coming at the bottom or at the
top end of the average spectrum. The strong attractive $L=0$
pairing of identical particles amplifies this effect and leads in
reality to a 100\% probability of $J_{0}=0$. Similar regularities
are associated with odd-$A$ and odd-odd systems. We relied mostly
on the nuclear structure as the basic and the best studied object
of applications. However, the similar physics of random
interactions and chaotic spin coupling certainly plays a role in
other mesoscopic systems, such as atomic clusters, metallic
grains, quantum dots and quantum spin glasses.

The new avenue were opened for general studies of quantum chaos.
From extremes of random matrix theory based on the most general
canonical Gaussian ensembles and later on the two-body random
ensemble, we proceed to investigation of random interactions fully
compatible with exact symmetries of finite systems. The new
effects that enter the game now are correlations between the
classes of states of different symmetry being governed by the
common Hamiltonian with a relatively small number of random
parameters. These correlations bring in the traces of new order
and collectivity solely created by random interactions along with
exact symmetries.

The pioneering paper \cite{JBD} on the subject was concluded by
the words that their studies ``have barely scratched the surface
of possible questions". Now we know more and perhaps can say that
we started to penetrate that surface. Nevertheless the problem is
not entirely solved, and the future advances seem to be quite
exciting and promising.

\section{Acknowledgments}

The authors are indebted to B.A. Brown, M. Horoi, D. Mulhall and
J. Roebke for creative collaboration. Constructive discussions
with G.F. Bertsch, R. Bijker, P. Cejnar, D. Dean, V.V. Flambaum,
A, Frank, F.M. Izrailev, D. Kusnezov, T. Papenbrock, N.A.
Smirnova, O.P. Sushkov, P. Van Isacker, H.A. Weidenm\"{u}ller,
Y.M. Zhao and A.P. Zuker at various stages of the work are
gratefully acknowledged. We acknowledge support from the NSF,
Grants Nos. PHY-0070911 and PHY-0244453, and from the US
Department of Energy, Nuclear Physics Division, under Contract No.
W-31-109-ENG-38.

\end{document}